\documentclass[a4paper,fleqn,usenatbib]{mnras}

\usepackage{newtxtext,newtxmath}

\usepackage[T1]{fontenc}
\usepackage{ae,aecompl}

\usepackage{graphicx}	
\usepackage{amsmath}	
\usepackage{amssymb}	
\usepackage[xindy]{glossaries}
\glsdisablehyper
\usepackage{subcaption}
\usepackage{tabularx}
\usepackage{xcolor}
\usepackage[normalem]{ulem}  

\newacronym{alfa}{ALFA}{Arecibo L-Band Feed Array}
\newacronym{dm}{DM}{Dispersion Measure}
\newacronym{frb}{FRB}{Fast Radio Burst}
\newacronym{fwhm}{FWHM}{Full-Width at Half-Maximum}
\newacronym{gbt}{GBT}{Greenbank Telescope}
\newacronym{htru}{HTRU}{High-Time Resolution Universe}
\newacronym{if}{IF}{Intermediate Frequency}
\newacronym{igm}{IGM}{Intergalactic Medium}
\newacronym{ism}{ISM}{Interstellar Medium}
\newacronym{ligo}{LIGO}{the Laser Interferometer Gravitational-Wave Observatory}
\newacronym{lo}{LO}{Local Oscillator}
\newacronym{nip}{NIP}{Non-image Processing}
\newacronym{pll}{PLL}{Phased-locked Loop}
\newacronym{rfi}{RFI}{Radio-frequency Interference}
\newacronym{rrat}{RRAT}{Rotating Radio Transient}
\newacronym{rm}{RM}{Rotation Measure}
\newacronym{ska}{SKA}{Square Kilometre Array}
\newacronym{sefd}{SEFD}{System Equivalent Flux Density}
\newacronym{snr}{S/N}{Signal-to-Noise Ratio}
\newacronym{sps}{SPS}{Single Pulse Search}
\newacronym{tab}{TAB}{Tied-Array Beam}
\newacronym{vlbi}{VLBI}{Very-Long Baseline Interferometry}
\newacronym{xao}{XAO}{Xinjiang Astronomical Observatory}

\newcommand{\SWIN}{Centre for Astrophysics \& Supercomputing, Swinburne University of Technology, Hawthorn, VIC 3122, Australia}

\title[FRB Verification]{Verifying and Reporting Fast Radio Bursts}

\author[G. Foster et al.]{Griffin Foster$^{1,2}$\thanks{E-mail: griffin.foster@physics.ox.ac.uk},
Aris Karastergiou$^{1,3,4}$,
Marisa Geyer$^{5}$,
Mayuresh Surnis$^{6,7}$,
\and
Golnoosh Golpayegani$^{6,7}$,
Kejia Lee$^{8}$,
Duncan Lorimer$^{6,7}$,
Danny C. Price$^{9,2}$,
\and
Kaustubh Rajwade$^{10}$
\\
$^{1}$University of Oxford, Sub-Department of Astrophysics, Denys Wilkinson Building, Keble Road, Oxford, OX1 3RH, United Kingdom\\
$^{2}$Department of Astronomy, University of California, Berkeley, 501 Campbell
Hall \#3411, Berkeley, CA, 94720, USA\\
$^{3}$Department of Physics and Electronics, Rhodes University,
    PO Box 94, Grahamstown 6140, South Africa\\
$^{4}$Physics Department, University of the Western Cape,
    Cape Town 7535, South Africa\\
$^{5}$SKA-SA, 3rd Floor, The Park, Park Road, Pinelands, Cape Town 7405, South Africa\\
$^{6}$Department of Physics and Astronomy, West Virginia University, Morgantown, WV 26505, USA\\
$^{7}$Center for Gravitational Waves and Cosmology, West Virginia University, Chestnut Ridge Research Building, Morgantown,\\ WV 26505, USA\\
$^{8}$Department of Astronomy \& Kavli Institute for Astronomy and Astrophysics, Peking University\\
$^{9}$\SWIN \\
$^{10}$Jodrell Bank Centre for Astrophysics, University of Manchester, Oxford Road, Manchester M13 9PL, United Kingdom\\
}

\date{Accepted XXX. Received YYY; in original ZZZ}

\pubyear{2018}

\begin{document}
\label{firstpage}
\pagerange{\pageref{firstpage}--\pageref{lastpage}}
\maketitle

\begin{abstract}
Fast Radio Bursts (FRBs) are a class of short-duration transients at radio
wavelengths with inferred astrophysical origin.  The prototypical FRB is a
broadband signal that occurs over the extent of the receiver frequency range, is
narrow in time, and is highly dispersed, following a $\nu^{-2}$ relation.
However, some FRBs appear band-limited, and show apparent scintillation, complex
frequency-dependent structure, or multi-component pulse shapes.  While there is
sufficient evidence that FRBs are indeed astrophysical, their one-off nature
necessitates extra scrutiny when reporting a detection as \emph{bona~fide} and
not a false positive. Currently, there is no formal validation framework for
FRBs, rather a set of community practices. In this article, we discuss potential
sources of false positives, and suggest a framework in which FRB-like events can
be evaluated as real or otherwise.  We present examples of false-positive events
in data from the Arecibo, LOFAR, and Nanshan telescopes, which while FRB-like,
are found to be due to instrumental variations, noise, and radio-frequency
interference.  Differentiating these false-positive detections from
astrophysical events requires knowledge and tests beyond thresholded
single-pulse detection.  We discuss post-detection analyses, verification tests,
and data sets which should be provided when reporting an FRB detection.
\end{abstract}



\section{Introduction}
\label{sec:intro}

\begin{figure*}
    \centering
    \begin{subfigure}[t]{0.45\textwidth}
        \centering\captionsetup{width=.95\linewidth}
        \includegraphics[width=1.0\textwidth]{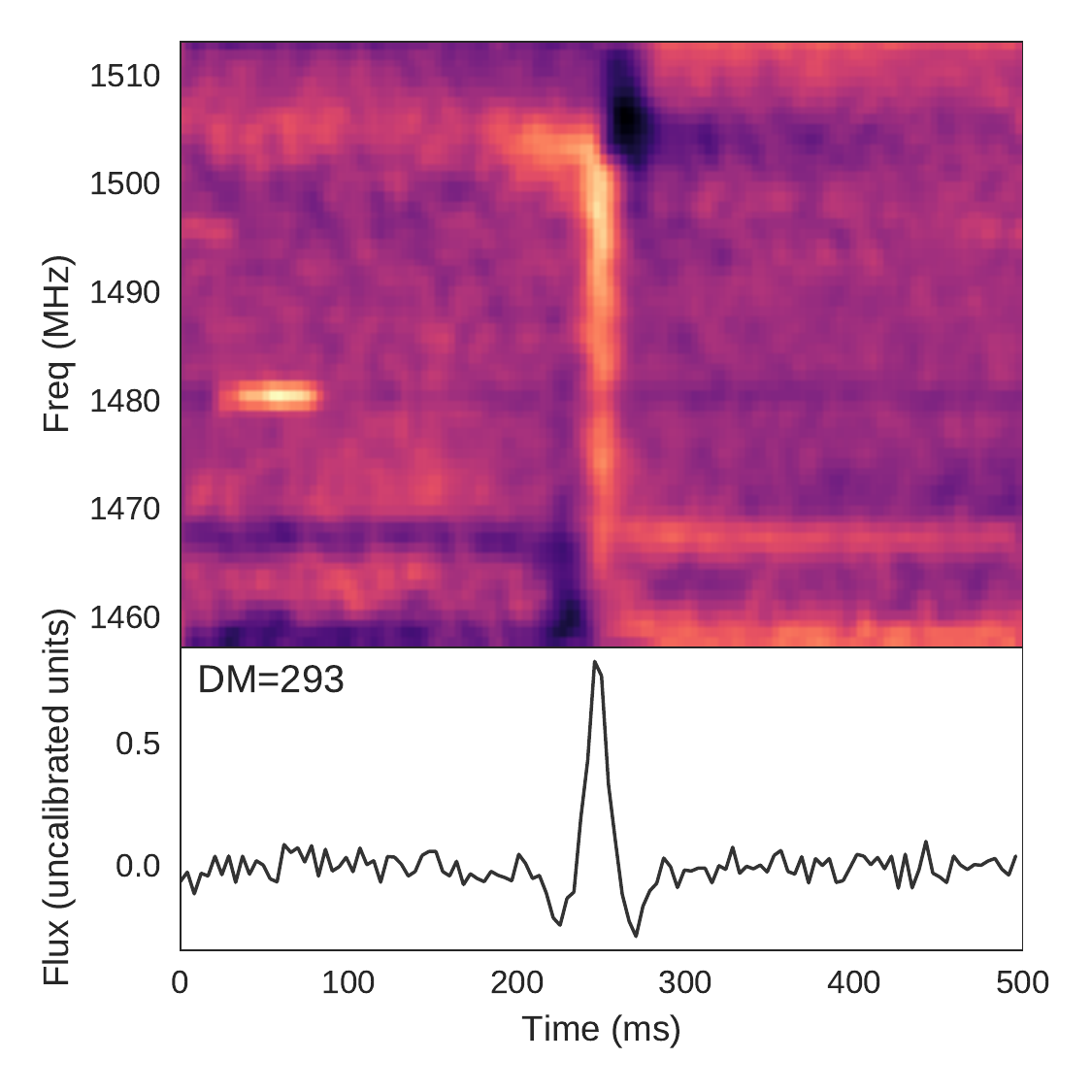}
        \caption{}
        \label{fig:beam0_dynamic_spec}
    \end{subfigure}
    \begin{subfigure}[t]{0.45\textwidth}
        \centering\captionsetup{width=.95\linewidth}
        \includegraphics[width=1.0\textwidth]{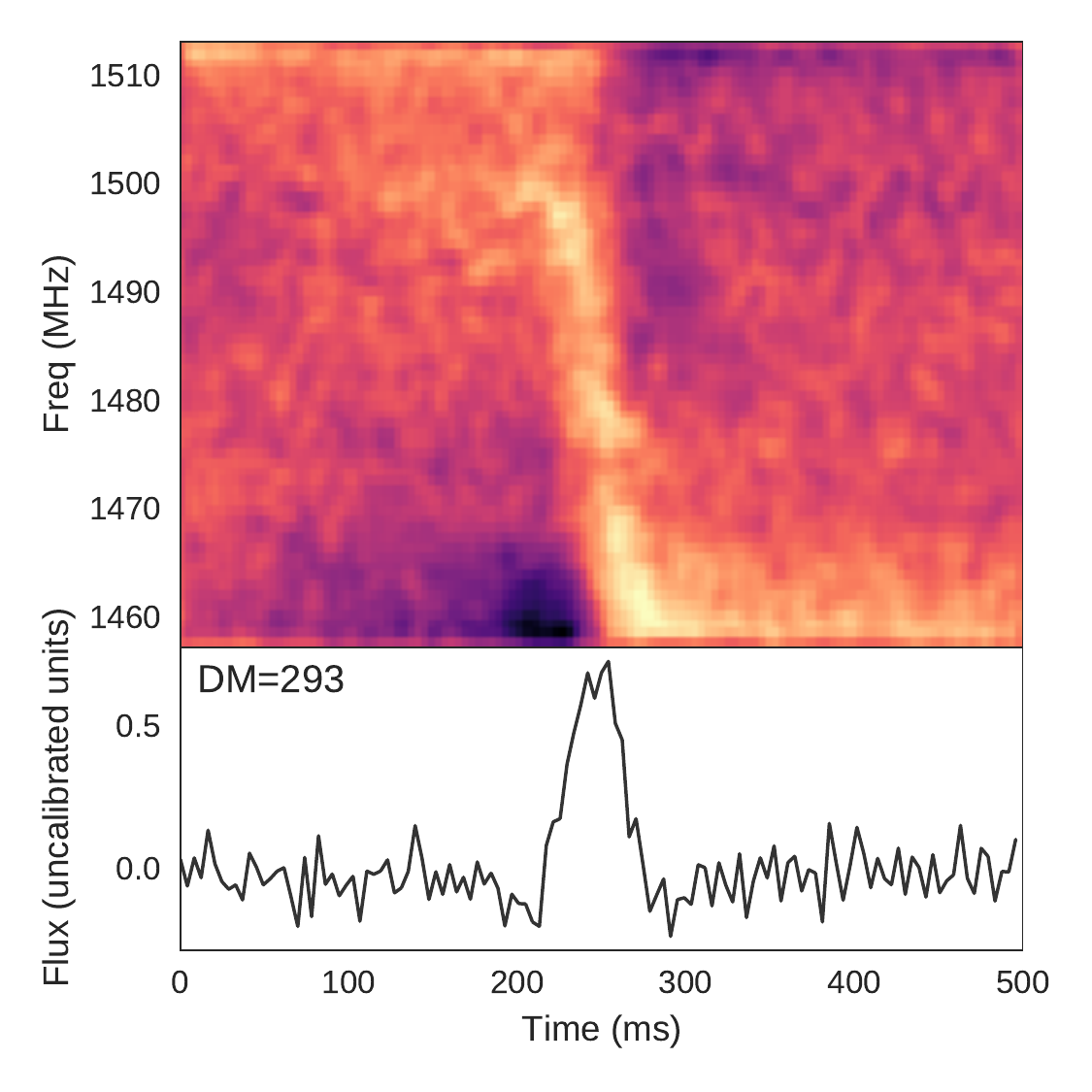}
        \caption{}
        \label{fig:beam5_dynamic_spec}
    \end{subfigure}
    \caption{
    Dynamic spectrum (top) and \gls{snr}-maximized de-dispersed time series
    (bottom) of an FRB-like event that was detected simultaneously in Beam 0 and
    5 of the ALFA receiver on December 4, 2016. The dynamic spectrum has been
    bandpass normalized. (a) Detected FRB-like event in Beam 0 of ALFA. The
    characteristic dip before and after the event is due to zero-DM removal
    which is part of the ALFABURST RFI exciser. The strong, narrow band source
    at 1480~MHz around 100~ms is due to a local RFI source. (b) Same event
    detected in Beam 5.
    }
    \label{fig:dynamic_spec}
\end{figure*}

The astrophysical origin of \glspl{frb} has been a mystery since they were
first reported \citep{2007Sci...318..777L}.  Though, the detection of multiple
\glspl{frb} in \cite{2013Sci...341...53T} put forth a convincing case for their
astrophysical nature and, subsequent detections have re-enforced this case.  The
\gls{dm} associated with the reported events indicate they occur well beyond our
Galaxy, possibly at cosmological distances.  Given their extragalactic distance,
observed fluxes suggest extremely energetic progenitors; however, their emission
mechanism remains unknown.  The consensus that \gls{frb} are astrophysical
events developed from  detections with multiple telescopes, over multiple
wavelengths, using different receiver systems.  \glspl{frb} are difficult to
detect as they require high-gain telescopes which typically have a small field
of view, such that despite many thousands of observing hours, only a few dozen
have been reported as of this writing
\citep{2016PASA...33...45P}\footnote{http://frbcat.org/}.

The prototypical \gls{frb} is broad-band across the observable bandwidth of the
receivers used. The pulse is narrow-in-time---on the order of a few milliseconds
in width---and highly dispersed, exhibiting a $\nu^{-2}$ relation with
frequency.  Despite follow-up observations in the direction of FRB events, only
FRB\,~121102 has been shown to repeat \citep{2016Natur.531..202S}. Several
reported detections deviate from this prototypical form, exhibiting complex
frequency-dependent structure and are possibly band-limited. The pulse width
also varies, due to either the emission process or propagation effects such as
scattering from the \gls{ism} and \gls{igm}.

These rare events are detected by automated, high-performance software pipelines
that extensively search a broad range of trial \glspl{dm} and pulse widths
\citep{Barsdell2012, 2015MNRAS.452.1254K, Bannister2017, Chime2018}. Each
de-dispersed time series is then thresholded -- any peaks above a minimum
\gls{snr} are reported as candidates. The number of candidates is usually
overwhelming due to \gls{rfi} and system gain variations. Initially, candidates
were reviewed manually; however, with the amount of data acquired in recent
surveys, it has become a significant time effort to do so. As such, and as our
understanding of the expected signal properties has grown, so too has the use of
machine-learning-based classifiers of candidate events
\citep[e.g.][]{Wagstaff2016, 2018MNRAS.474.3847F, 2018MNRAS.478.1209F, Connor2018}.

As FRBs are rare and appear not to repeat (with the notable exception of
FRB\,121102), being able to confidently verify them is an important issue. There
is significant \gls{rfi} detectable at all radio observatories, and there are
known anthropogenic sources that appear FRB-like \citep{2015MNRAS.451.3933P}.
Given the significant number of \gls{dm} trials and high-time resolution of the
spectra in a typical survey, there are a large number of false-positives (type-I
errors) that pass the automated post-processing detection thresholds.  This is
by design, as we would like to severely limit the potential for false-negatives
(type-II errors) in our detection pipelines by accepting a number of
false-positives during automated searches and manually discarding them later.
But, given the large sampling of the parameter search space it can be difficult
to differentiate between a true, astrophysical \gls{frb} (true-positive) and a
`terrestrial \gls{frb}' (false-positive) due to \gls{rfi}, systematics, or other
local effects. As the survey time increases, the likelihood of detecting such a
false-positive will increase. 

In this paper, we present a list of verification criteria to counter false
detection of `terrestrial \glspl{frb}', that can be applied \emph{post facto} to
recorded data. We firstly discuss examples of FRB-like sources, which after
investigation, were shown to be non-astrophysical (\S~\ref{sec:false-pos}).
Using these and previously reported \glspl{frb}, we develop a set of criteria to
test detections against (\S~\ref{sec:verify_crit}). We then apply these criteria
to some of the previously reported detections
(\S~\ref{sec:appln_to_previous_detections}) as examples of usage, then discuss
future observational methods to reduce ambiguity (\S~\ref{sec:future_methods}).
We conclude with suggestions as to what data, in addition to the detected
dynamic spectrum, should be reported to provide a robust statement about
\gls{frb} detections in general.

\section{False-Positive FRB Detections}
\label{sec:false-pos}

The rate of false-positive detections is set by the minimum \gls{snr} threshold,
the parameter search space, and the terrestrial environment of the observatory.
We use the term `terrestrial' to encompass multiple effects: anthropogenic radio
signals and variations in the observing system.

Most potential false-positive events are flagged using filters and classifier
models.  The remaining events are commonly examined by eye as expert human
knowledge appears to be the best way to classify a detection. However, time
constraints do not allow for all events to be classified in this way. On
inspection, true astrophysical events and terrestrial events can be difficult to
differentiate between, even by the expert eye.  This is not surprising, as they
have passed the detection pipeline thresholds.

Studies of the telescope state during detection, can often assist in determining
the origin of an event. Nonetheless, ambiguities can remain.  In this section we
present examples of such events---which on initial inspection appear
astrophysical, but after further investigation prove terrestrial in origin.

\subsection{ALFA Terrestrial FRB Candidate}
\label{sec:D20161204}

In the two years of the initial ALFABURST survey \citep{2017ApJS..228...21C,
2018MNRAS.474.3847F}, over 200,000 8.4-second data windows were recorded in which
the \gls{frb} search pipeline detected candidates using a minimum \gls{snr}
detection threshold of 10. The vast majority of these events were due to
\gls{rfi} and instrumental variations, while others were due to bright single
pulses of known pulsars.  The random forest-based classifier model outputs an
ordered list of events based on the likelihood of the event being a pulse.

A narrow-in-time, broad-in-frequency, millisecond pulse was detected with the
ALFABURST system at 09:31:06 Arecibo local time (UTC 13:31:06) on 2016, December
4 in Beam 0 (the central beam) of the \gls{alfa}
receiver\footnote{http://www.naic.edu/alfa/} (Figure
\ref{fig:beam0_dynamic_spec}) which we label as the D20161204 event for this
discussion. ALFABURST was processing 56~MHz of bandwidth between 1457~MHz and
1513~MHz. The \gls{snr} of this pulse is maximized when the pulse is
de-dispersed with a \gls{dm} of 293~pc~cm$^{-3}$, and the native 256~$\mu$s time
resolution is decimated by a factor of 16 to a time resolution of 4.096~ms. The
de-dispersed time series shows an approximately 20~ms \gls{fwhm} pulse.  The dip
before and after the pulse is due to the zero-DM filtering (i.e. the moving
channel average is subtracted) before pulse detection
\citep{2009MNRAS.395..410E}. This is a simple way to remove a drifting gain
baseline at the cost of removing some of the overall pulse power, particularly
at low DM trials.

On initial inspection, this event appears astrophysical. From radiometer noise
considerations, the flux density of the pulse is
\begin{equation}
S = \textrm{SEFD} \frac{\textrm{S/N}}{\sqrt{D \; \Delta \tau \;
\Delta \nu}}
\end{equation}
where SEFD is the system-equivalent flux density of the telescope, $\Delta \tau$
is the time resolution of the spectra, $\Delta \nu$ is the total frequency
bandwidth of the spectra, and $D$ is the time decimation factor which maximizes
the \gls{snr} of the detections.  Adopting a \gls{sefd} of 3~Jy for the
\gls{alfa} receiver\footnote{http://www.naic.edu/alfa/},  results in \mbox{$S =
66$}~mJy from beam 0, which would be lower than what has been observed for any
previously detected \glspl{frb} \citep{2016PASA...33...45P}.  However, this flux
estimate is a lower limit, assuming the source is at the center of the beam. The
pulse width is large for an \gls{frb} but within the range of those previously
reported.

As ALFA is a multi-beam system, we checked if the event occurred in any other
beams. Of the six other \gls{alfa} beams, the event also appears in beam 5
(Figure \ref{fig:beam5_dynamic_spec}), which is adjacent to beam 0.  This pulse
lines up with the beam 0 event in time exactly, but the event \gls{snr} was
maximized ($S/N=16$) for a DM of 829~pc~cm$^{-3}$, but this higher \gls{dm} is
due to a slope in the bandpass. Flattening the bandpass, and re-performing the
DM-trial fit, resulted in a \gls{snr}-maximized \gls{dm} closer to that of beam
0.  Our \gls{rfi} clipping method was applied during this event which is known
to affect the pulse shape, and generally results in a maximized \gls{snr} at a
higher DM trial than expected.  We thus conclude that the Beam 0 and Beam 5
event are the same event.

\begin{figure}
    \centering
    \begin{subfigure}[t]{0.5\textwidth}
        \centering\captionsetup{width=.95\linewidth}
        \includegraphics[width=1.0\textwidth]{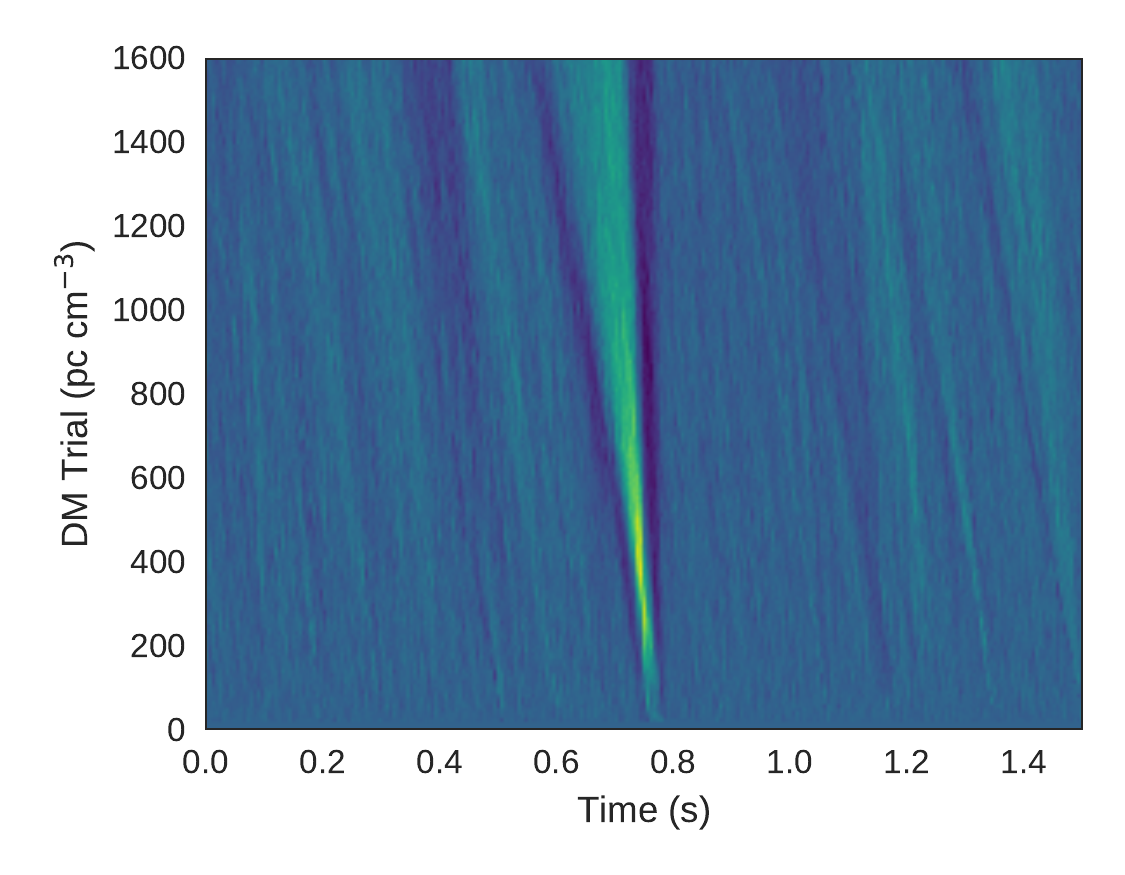}
        \caption{DM vs. time plot of for a 1.5~second window centred on the
        December 4th event in Beam 0. The \gls{snr} peaks at a DM of
        293~pc~cm$^{-3}$. There is a significant detection at larger DM trials
        due to the width of the pulse.
        }
        \label{fig:dm_time_event}
    \end{subfigure}
    \begin{subfigure}[t]{0.5\textwidth}
        \centering\captionsetup{width=.95\linewidth}
        \includegraphics[width=1.0\textwidth]{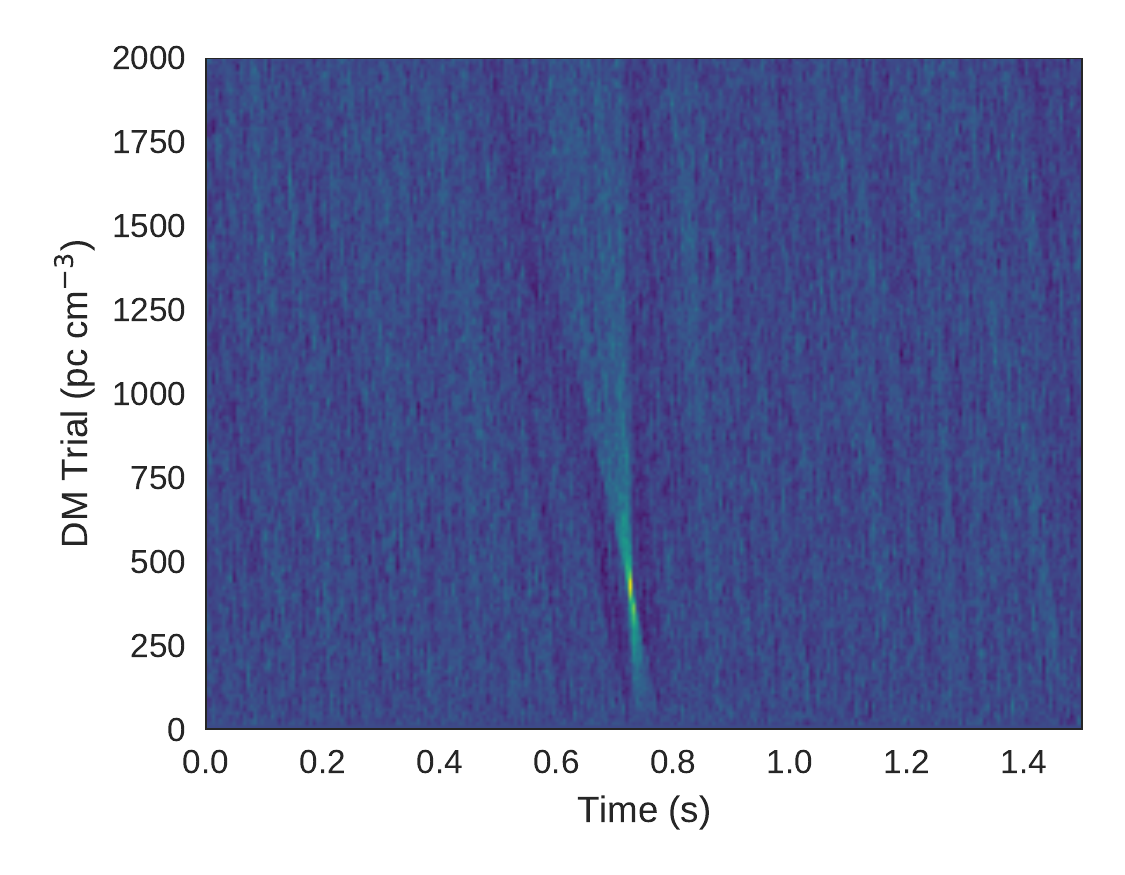}
        \caption{DM vs. time plot of a bright single pulse from PSR B1859+03
        which has a DM of 402~pc~cm$^{-3}$ and pulse width of 11~ms (W50).
        }
        \label{fig:dm_time_B1859}
    \end{subfigure}
    \caption{DM-space plot shows the characteristic butterfly pattern of the
    narrow-in-time, dispersed pulse detected by ALFABURST at a different epoch.
    A single pulse detection of PSR B1859+03 is shown for reference.
    }
    \label{fig:dm_time}
\end{figure}

We checked if any other events occurred with in a larger time window of the
detection, as the event could be a high \gls{snr} detection of a series of
\gls{rfi} events.  In the immediate period before and after the pulse, there are
no similar events (Figure \ref{fig:dm_time}). The event appears to be isolated
in time, with a fairly compact representation in DM-time space (Figure
\ref{fig:dm_time_event}) similar to that of a single pulse from a high DM
pulsar, such as PSR B1859+03 (Figure \ref{fig:dm_time_B1859}).

As ALFABURST is a commensal observing program we checked the telescope pointing,
state, and receiver stability.  The telescope pointing was fixed in azimuth and
elevation, corresponding to a fixed declination (+15:11) and drifting in RA
(event detected at RA=14:42). No known pulsar or \gls{rrat} with a similar DM
lies within the beam primary lobe of this pointing.

The telescope logs provide the first indication that this event is spurious. As
ALFABURST is a commensal observation survey the search pipeline regularly checks
the status of the receiver turret and the centre observing frequency. \gls{alfa}
was logged as active and the centre observing frequency remained unchanged
during this time, but the turret was in a different position than where ALFA
should be if it were at the secondary focus.

The observing schedule for the morning of December~4, was project
P3080\footnote{http://www.naic.edu/vscience/schedule/tpfiles/MichillitagP3080tp.pdf}
using \gls{alfa} to perform an \gls{frb} survey of the Virgo cluster until 09:00
local time, followed by Project
R3037\footnote{http://www.naic.edu/vscience/schedule/tpfiles/TaylortagR3037tp.pdf},
an S-Band radar observation.  The event occurred during a time of no active
observation with \gls{alfa} active but in the wrong turret position. Thus,
\gls{alfa} was at a significantly reduced sensitivity than expected when in the
correct position.

\begin{figure}
    \includegraphics[width=1.0\linewidth]{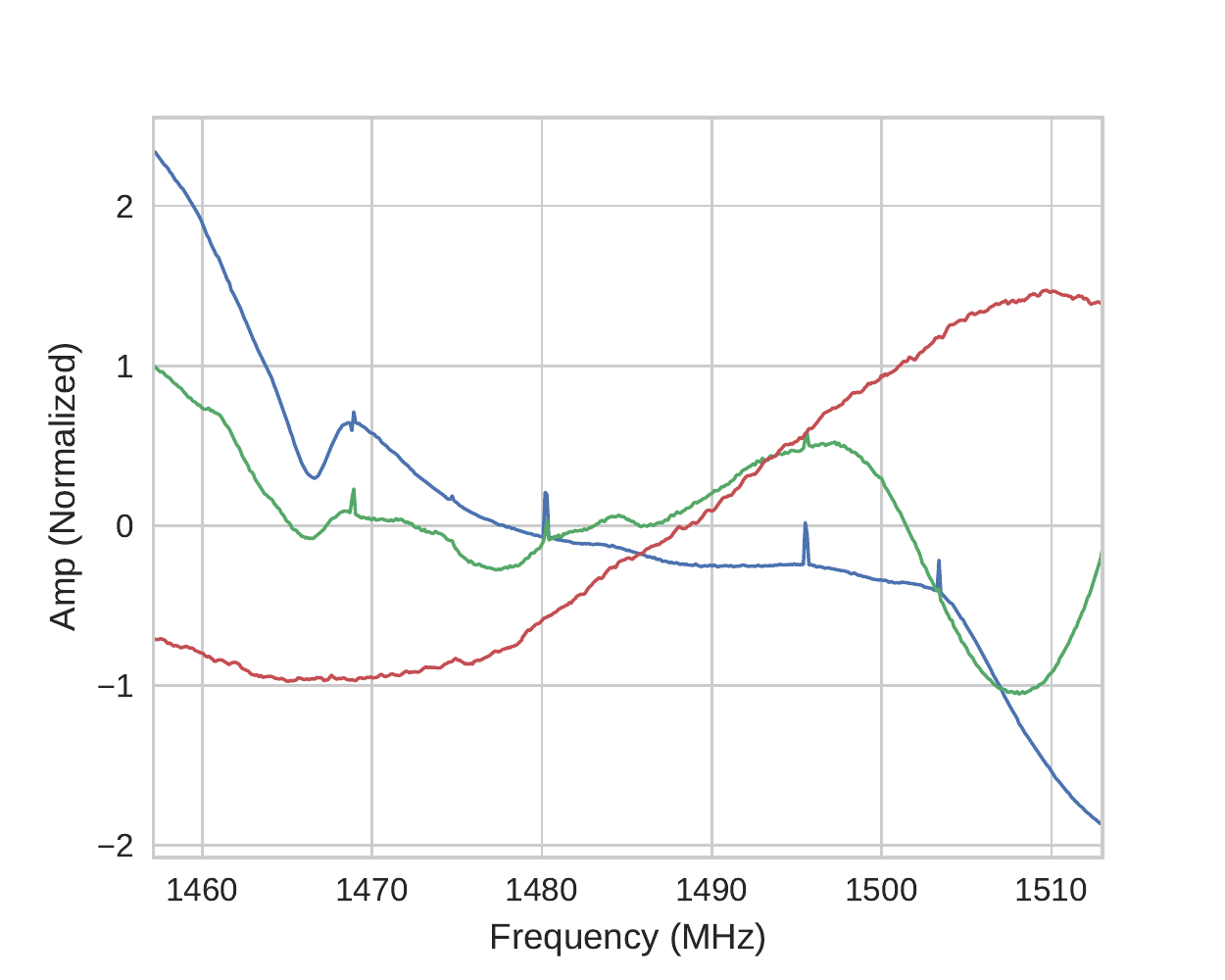}
    \caption{Average bandpass response during the December 4, 2016 event for
    Beam 0 (green) and Beam 5 (blue). A typical bandpass (red) is plotted for
    reference. The narrow features at 1468, 1480, 1496, and 1504~MHz is due to
    local \gls{rfi}. These bandpasses have been normalized in the detection
    pipeline.
    }
    \label{fig:bandpass_response}
\end{figure}

As the telescope was in an inactive observing state we checked if the
electronics were operating as expected during normal ALFA observations.  The
average bandpass of Beam 0 and Beam 5 during the time of the event shows that
the shape and system noise appear atypical compared to normal observations
(Figure \ref{fig:bandpass_response}).  Beam 0 and 5 bandpasses appear similar
and have overlapping narrow-band features at 1468, 1480, 1496, and 1504~MHz due
to the settling of \gls{pll} in a local oscillator associated with an on-site
\gls{rfi} monitor.  The system noise appears higher during the event, which
leads to smoother bandpasses than typical.  In the detection pipeline the data
are normalized, which removes all absolute scaling. This indicates that the
\gls{sefd} is too low in our flux calibration.  This increase in system noise is
due to the change in turret position, causing the \gls{alfa} feed to pick up
reflections from other equipment in the dome, and the dome itself as a warm
source.  This event is most likely due to the analogue electronics becoming
non-linear from this increase in system noise.

\begin{figure*}
    \includegraphics[width=1.0\linewidth]{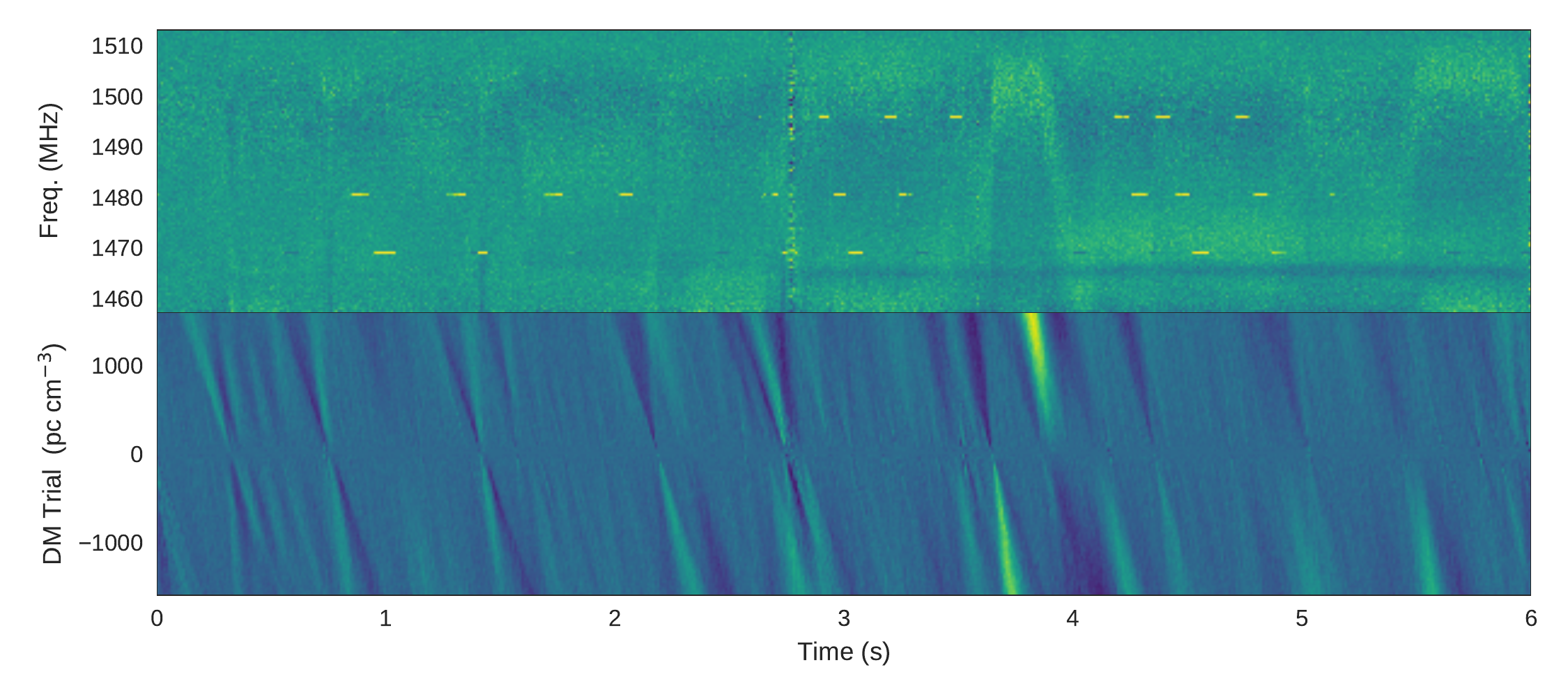}
    \caption{
    Dynamic spectrum (top) and DM-time plot (bottom) of 6 seconds from beam
    0 approximately 80 seconds before the D20161204 event.  The dynamic spectrum
    shows frequency evolution of the bandpass as a function of time (e.g. the
    sinusoidal shape around the four second mark) results in significant
    detections across the trial DM range -1600 to 1600~pc~cm$^{-3}$. The
    narrow-band \gls{rfi} at 1467~MHz is due to a local \gls{rfi} source. The
    broad-band pulse near the three second mark is an \gls{rfi} source which has
    been partially removed.
    }
    \label{fig:beam0_80s}
\end{figure*}

Evidence for analogue gain variations is further justified by considering the
data over a larger time window.  Approximately 80 seconds before the event,
frequency-varying structure across the band (top of Figure \ref{fig:beam0_80s})
was present.  Though not as narrow-in-time as the event, they appear related to
the same phenomenon.  The DM$-$time plot (bottom of Figure \ref{fig:beam0_80s})
shows that much of the structure would be detected as dispersed pulses.  In
particular, at around the four second mark the structure would result in a
wide-in-time, highly dispersed pulse detection.  The structure immediately
proceeding it would be detected as a negatively dispersed pulse, but only
positive \gls{dm} candidates are recorded.  The narrow-in-frequency, repeating
feature present at 1467~MHz is the same feature as seen in the time-averaged
bandpass (Figure~\ref{fig:bandpass_response}).

In isolation, and taking into account the one-off, transient nature of FRBs, the
initial beam 0 detection reasonably appears astrophysical. But, after further
checks on the telescope state and examining data across a larger time window
(several seconds) around the event, we can confidently classify this event as
terrestrial. This has motivated us to develop a set of tests to use in verifying
an event as astrophysical, this is discussed in detail in Section
\ref{sec:verify_crit}.

\subsection{Low-S/N False-Positive Detections}
\label{sec:low_snr}

\begin{figure}
    \includegraphics[width=1.0\linewidth]{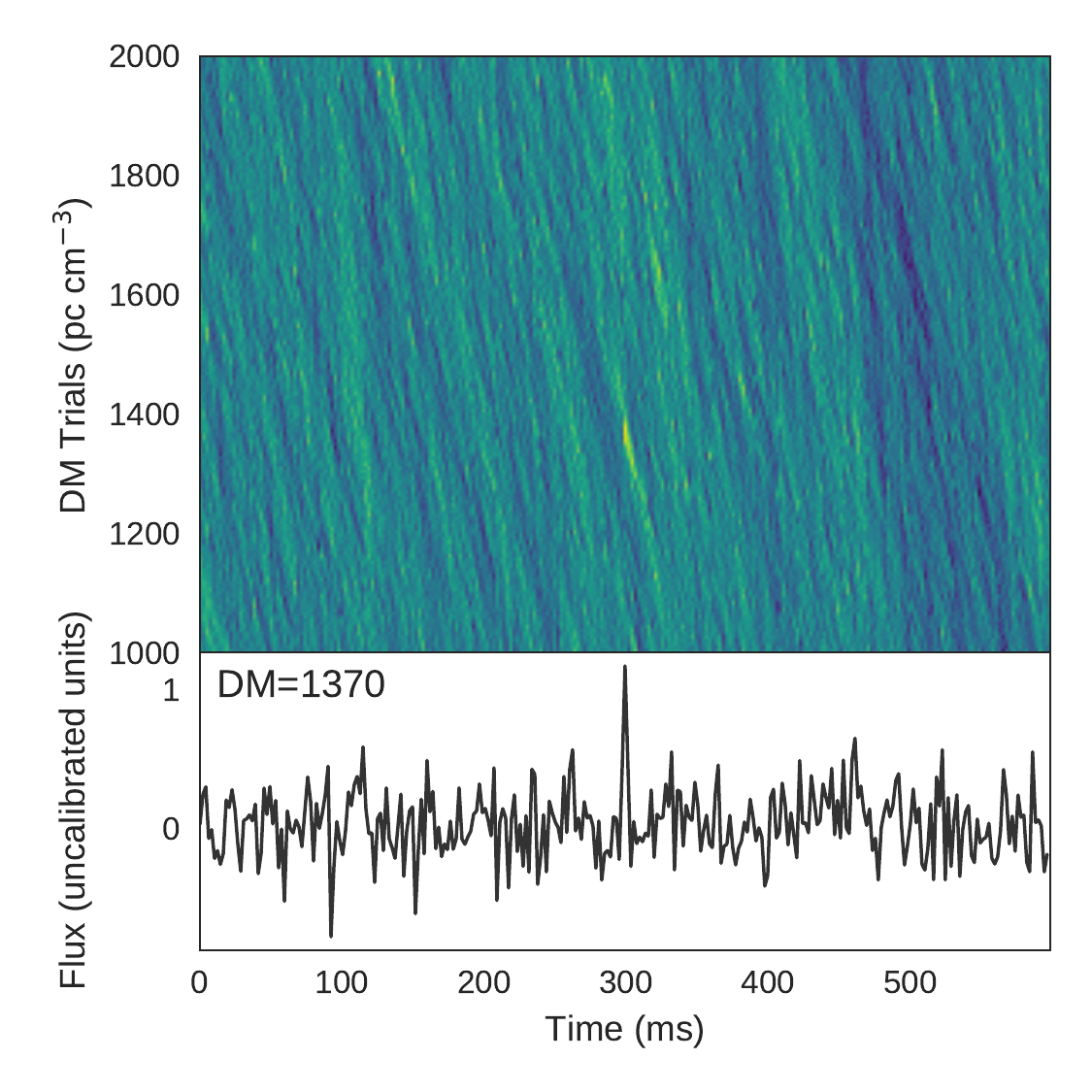}
    \caption{DM-space plot and time series of a high DM event with reported
    \gls{snr} above 10-$\sigma$ on July 30, 2015. After re-analysis and review
    of the telescope meta-data it was determined that this event was due to
    local system noise.
    }
    \label{fig:D20150730}
\end{figure}

While the D20161204 event (Section \ref{sec:D20161204}) is a rare false-positive
detection based on a unique telescope state, low-\gls{snr} false-positive
candidates occur regularly\footnote{A minimum \gls{snr} cut-off (usually 6--10)
is set to limit their occurrence.}. For example, on July 30, 2015 an apparently
10-$\sigma$ event with a DM of 1370~pc~cm$^{-3}$ was detected (Figure
\ref{fig:D20150730}) by the ALFABURST pipeline in beam 5. The pulse can barely
be seen in the dynamic spectrum, but in the DM-time plot, there is a compact
peak centered at a DM of 1370~pc~cm$^{-3}$ with no other similar events
apparent.  After re-analysis of the dynamic spectrum noise, this event only has
an \gls{snr}$\sim$6, but was reported as a higher \gls{snr} because of the data
window size used to calculate the system noise.  Locally in time, there is an
increase in system noise, leading to a reduced \gls{snr}.  This is not an
uncommon event and many of these events have no immediate explanation.

These types of events prove to be difficult to validate as either astrophysical
or terrestrial. The choice of minimum \gls{snr} is partially set by the
willingness of the observers to sort through false-positive events.  There are
likely many low-\gls{snr} \glspl{frb} that are either not detected at all or
labelled as false-positive events.  The reported detection of a pulse from
FRB\,121102 with APERTIF \citep{atel10693} had an \gls{snr}$\approx 4$. In a blind
survey across different sky positions and trial \glspl{dm}, this would not be a
significant detection.  But, the sky position and \gls{snr}-maximized DM of the
repeating FRB\,121102 is known, thus a lower \gls{snr} detection may be
sufficient.

\subsection{ARTEMIS Radar Detection}
\label{sec:LOFAR_RADAR}

ARTEMIS \citep{2015MNRAS.452.1254K} is a low-frequency \gls{frb} survey using
the Rawlings Array, the LOFAR-UK station at Chilbolton Observatory.  The survey
uses a similar fractional bandwidth ($\sim 0.04$) to ALFABURST, but is centered
at 145~MHz.  In this survey, known pulsars regularly transit the beams that are
fixed in azimuth and elevation, and single pulses are routinely observed.  An
\gls{rfi} excision algorithm has been developed for this survey that
successfully removes the majority of false-positive events. Over many thousands
of observing hours, rare events occasionally pass this filter.

\begin{figure}
    \includegraphics[width=1.0\linewidth]{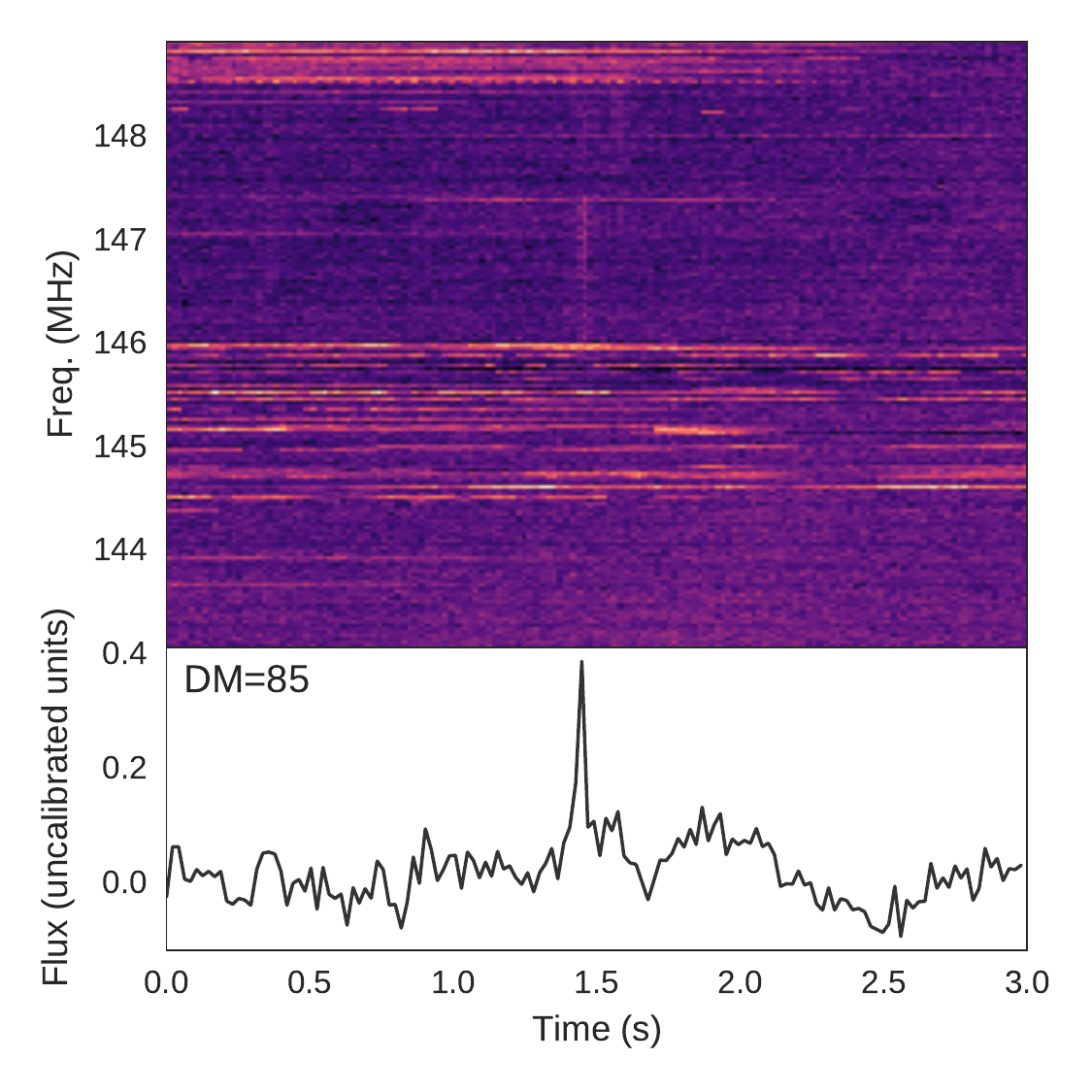}
    \caption{A dispersed pulse detected by the automated ARTEMIS search pipeline
    at the LOFAR-UK station. The \gls{snr} is maximized when the signal is
    dedispersed to a DM of 85~pc~cm$^{-3}$. The dynamic spectrum has a time
    resolution of 20~ms and frequency resolution of 3~kHz.
    }
    \label{fig:lofar_dynamic}
\end{figure}

A particularly interesting event for which the \gls{snr} was maximized at a
\gls{dm} of 85~pc~cm$^{-3}$ is shown in Figure \ref{fig:lofar_dynamic}. Though
this event has a low DM compared to the reported \glspl{frb}, this still proves
to be a relevant example as discussed later in this section.  The narrow-in-time
pulse can be seen in the dynamic spectrum at frequencies above 146 MHz, but not
at lower frequencies where it could be hidden by narrow band \gls{rfi}.  The
de-dispersed time series shows a high-\gls{snr} detection of a pulse of
approximately 20~ms in width.  The beam pointing during the time of the event is
not associated with any known pulsar or RRAT around a DM$\sim$85~pc~cm$^{-3}$.

\begin{figure}
    \includegraphics[width=1.0\linewidth]{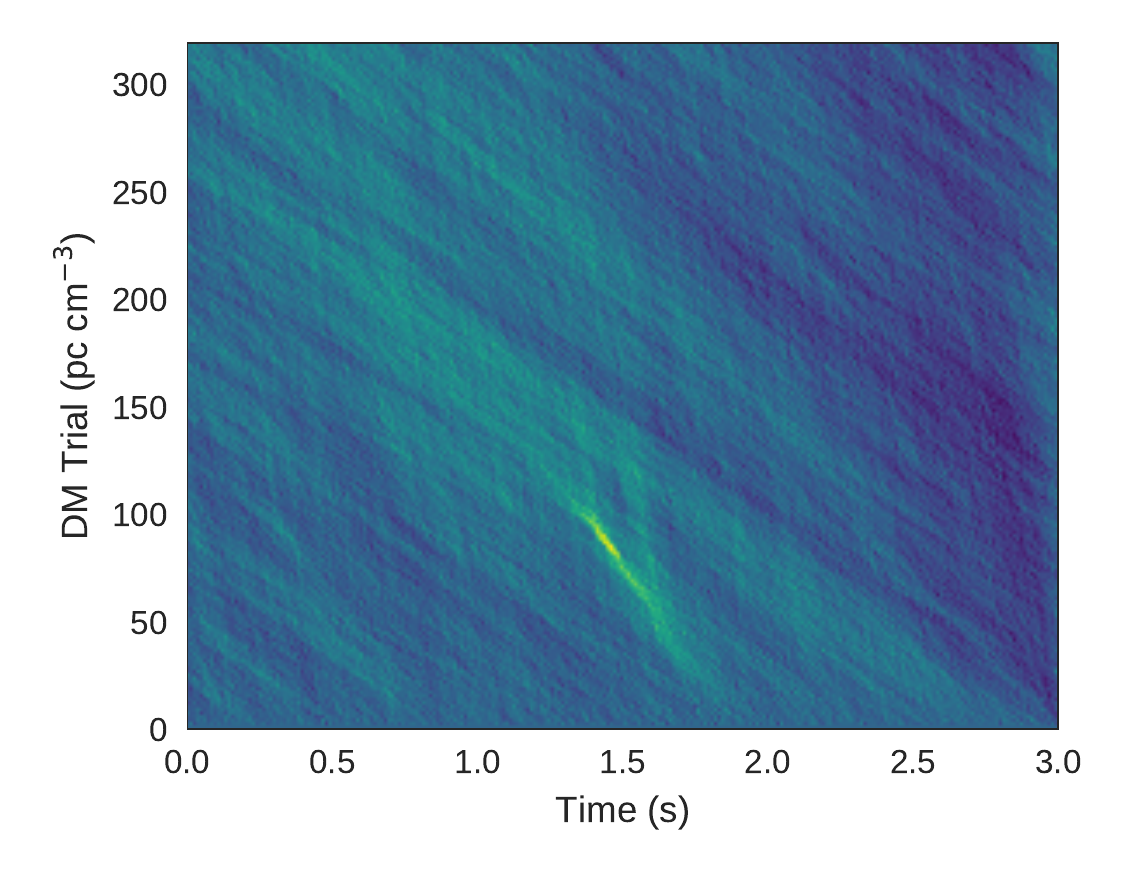}
    \caption{DM-space plot over the DM trial range of the ARTEMIS survey. A
    strong, compact detection occurs at a DM of 85~pc~cm$^{-3}$ with no other
    apparent events during the time.
    }
    \label{fig:lofar_dm_time}
\end{figure}

Plotting the event in DM-time space across the ARTEMIS DM range
($0-320$~pc~cm$^{-3}$, Figure \ref{fig:lofar_dm_time}) shows a strong, compact
detection as expected of a dispersed pulse. But, the apparent limited bandwidth
of the pulse means that the pulse does not necessarily follow a $\nu^{-2}$
dispersion relation.

\begin{figure}
    \includegraphics[width=1.0\linewidth]{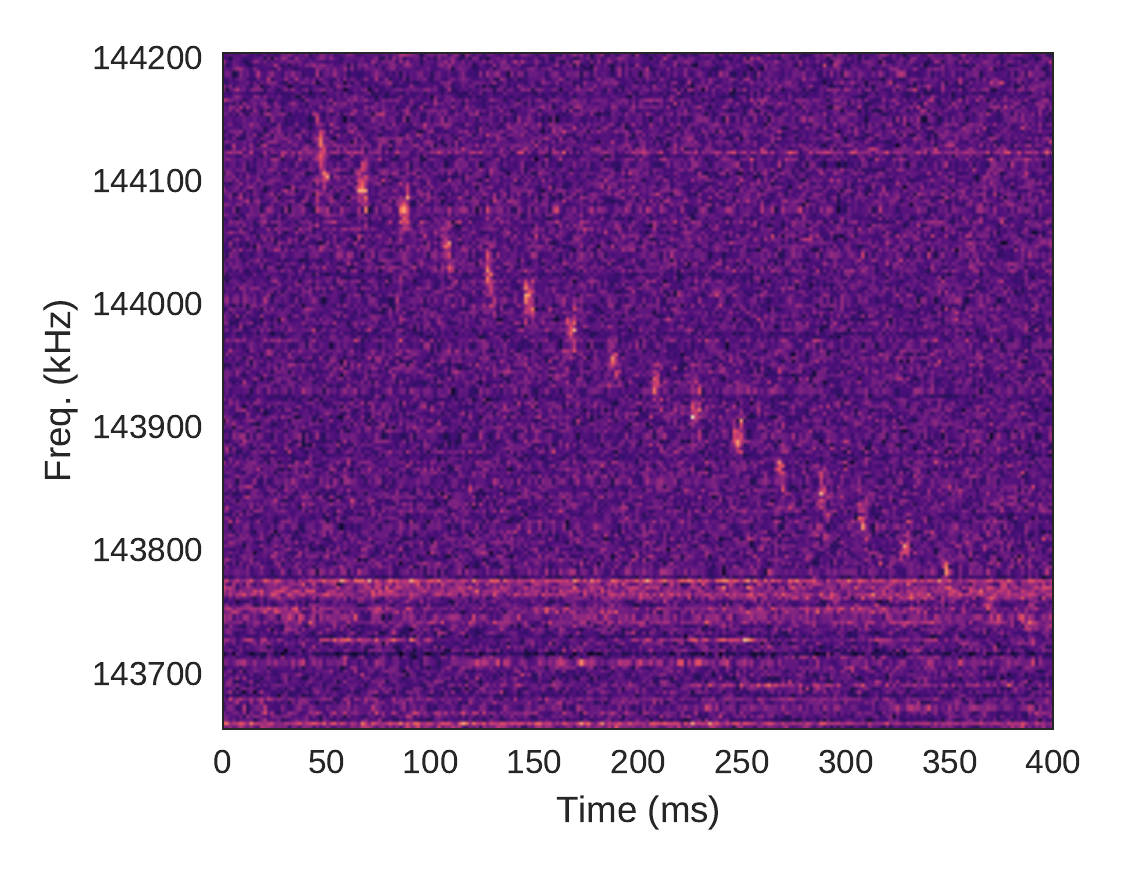}
    \caption{A zoomed in view of the event in Figure \ref{fig:lofar_dynamic} at
    high time (1~ms) and frequency (1.5~kHz) resolution shows the distinct
    pattern of a linear frequency-modulated radar pulse.
    }
    \label{fig:lofar_dynamic_high}
\end{figure}

The ARTEMIS search pipeline, like most \gls{frb} search pipelines, decimates the
dynamic spectrum in time to search over a range of pulse widths. The \gls{snr}
of the event shown in Figure \ref{fig:lofar_dynamic} is maximized for a time
decimation factor of 64. With a native resolution of $327.68~\mu s$, this
results in a decimated time resolution of 20~ms. At this resolution, the pulse
appears to be a continuous broadband pulse. When an event is detected in the
pipeline, the dynamic spectrum is saved at the original resolution. Plotting at
1~ms time resolution (Figure \ref{fig:lofar_dynamic_high}), the repeating nature
of a linear frequency-modulated signal used for pulse compression in radar can
be seen.

In radar signals, the bandwidth of the transmitter provides information on the
range and direction of a target. A narrow-band radar transmitter can be used to
approximate a dispersed, wide-band pulse by modulating the frequency of the
transmitted pulse. The narrow-band (in frequency) pulse is stepped in frequency
across a transmission band. Between each step the pulse is not transmitted,
resulting in the gaps in time between pulses, such as in
Figure~\ref{fig:lofar_dynamic_high}.  An increase in the delay between the
narrow-band pulses will result in a more dispersed wide-band pulse.

Linear frequency-modulation is the most typical form of chirp compression, but
non-linear methods are also used. Such a modulation technique may be the origin
of other detected signals (see next section).  There are a number of allocated
radar usages in the LOFAR observing band which could be the source of the
observed radar pulse \citep{ofcom2017}.  Radar is used from UHF to C-band
(300~MHz -- 8~GHz), covering a wide range of frequencies at which \gls{frb}
surveys operate. We could not determine the exact origin of the radar pulse,
since radar is used for commercial and military purposes, most of these signal
specifications, modulations, and source locations are proprietary.

As the radar signal can resemble a dispersed pulse, we expect to detect such
signals with \gls{frb} pipelines.  Verification of this event is straightforward
when the high-time and frequency resolution data are available to reveal the
narrow-band, pulsed nature of the event.

\subsection{XAO Repeating Event}
\label{sec:xao_event}

\begin{figure}
    \includegraphics[width=1.0\linewidth]{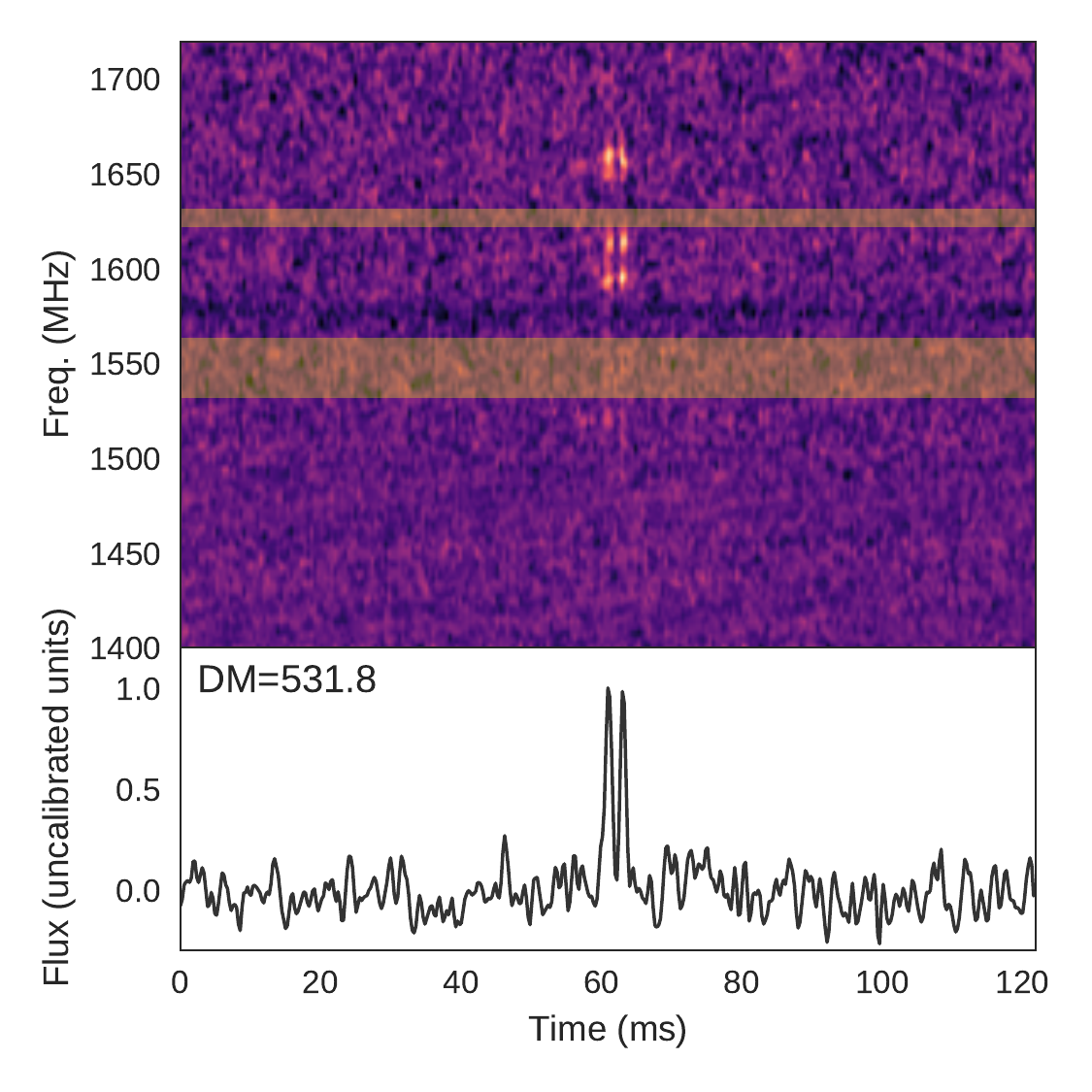}
    \caption{Example of a detected pulse with the Nanshan Radio Telescope at XAO
    which is \gls{snr} maximized at a \gls{dm} of $531.8$~pc~cm$^{-3}$.
    Hundreds of such pulses were detected over a period of a few hours. The
    orange bars represent regions that have had significant constant-in-time
    \gls{rfi} replaced by noise.
    }
    \label{fig:xao_dynamic}
\end{figure}

The 25-m Nanshan Telescope at \gls{xao} is currently running an FRB survey which
covers over 300~MHz at L-band (1 -- 2~GHz), sampling at $64 \; \mu s$
resolution. On November 18, 2016 hundreds of bright, dispersed pulses were
detected. The pulses varied in \gls{snr}, but had the same \gls{snr}-maximized
DM of 531.8~pc~cm$^{-3}$. The pulses show a distinct double peak (each peak
$\sim 2$~ms wide) separated by $\sim 3$~ms (Figure \ref{fig:xao_dynamic}). The
pulses are only apparent in a portion of the band.

A periodicity search revealed a periodicity of $\sim 1.7$~s, but the timing
residuals were orders of magnitude higher \mbox{($\sim 0.01$~s)} than that of a
typical pulsar periodicity. Further, the pulses were seen at different pointings
across the sky.  Most were detected at a low altitude pointing angle, but some
were detected close to zenith.  Dispersion in these pulses was found to deviate
from a $\nu^{-2}$ relation by $1.5 \sigma$. A dispersed astrophysical source
should always follow a cold plasma $\nu^{-2}$ relation. This rules out a
pulsar, or an astrophysical source in general.

A thorough search of possible local sources, such as new equipment, vehicles,
and aircraft was performed, with no obvious candidate being found. Detection of
multiple pulses at different beam pointings indicates the source may be directly
illuminating the feed.

The Nanshan L-band receiver is a single pixel system. A multi-beam system, such
as the Parkes 13-beam or ALFA 7-beam receivers, would likely have detected these
events in several beams, which would indicate a terrestrial origin to the event.

Had only a single pulse been detected, for example if the source was weaker, or
the telescope was only sensitive to the highest \gls{snr} event, it would be
difficult to show that the event was due to \gls{rfi}.  Multiple reported
\glspl{frb}, including the repeater FRB\,121102, do not cover the entire
observing band. This has been explained by various intrinsic or intermediate
effects, e.g.  scintillation \citep{2015Natur.528..523M,0004-637X-863-2-150} and
plasma lensing \citep{2017ApJ...842...35C}. In the case of a single detected
pulse, it would be reasonable to report it as an astrophysical \gls{frb}.

Since the initial detection of the pulses in November 2016 the pulses have not
been re-detected. While the source is terrestrial in origin, these pulses
present a situation where terrestrial sources can easily be misidentified as
one-off astrophysical events. This event, along with the other events presented
in this section should present a case for developing common criteria, observing
strategies, and data recording to improve the confidence in \gls{frb}
detections.

\subsection{Parkes Perytons}
\label{sec:perytons}

A class of non-astrophysical, yet FRB-like signals were found in high-time
resolution data around 1.4~GHz with the Parkes Telescope
\citep{2011ApJ...727...18B}. These signals, dubbed `Perytons', are dispersed in
frequency like \glspl{frb}, but do not obey the $\nu^{-2}$ cold plasma relation.
Initial reporting classified these signals as terrestrial as they occurred in the
near-field (i.e. they appeared in most of the beams of the Parkes Multi-beam
receiver at approximately equal \gls{snr}) indicating the source was local to
the telescope site.  Additional Perytons were detected while an on-site,
broad-band \gls{rfi} monitor recorded simultaneous \gls{rfi} at $2.4$~GHz
\citep{2015MNRAS.451.3933P}. Further analysis led to a time of day dependence
relation that indicated they were generated by equipment in use during normal
working hours.  Ultimately, these were found to be a form of \gls{rfi} generated
by microwave ovens.

\subsection{Simulated False-Positive Pulses}
\label{sec:sim_pulses}

As shown in the examples of this section, narrow-in-time pulses from terrestrial
origins can take multiple forms which follow an approximately $\nu^{-2}$
dispersion relation. This can be due to instrumentation variation effects, or as
anthropogenically created signals such as chirps with frequency-modulated
structure.  An \gls{frb} dedispersion pipeline acts as an approximate matched
filter to a subset of these signals leading them to be detected with high
\gls{snr}. By simulating a simple receiver and dispersed pulse model we can show
that \gls{frb} pipeline false-positive detections can be produced over a broad
range of \glspl{dm} and \glspl{snr}.

\begin{figure*}
    \includegraphics[width=1.0\linewidth]{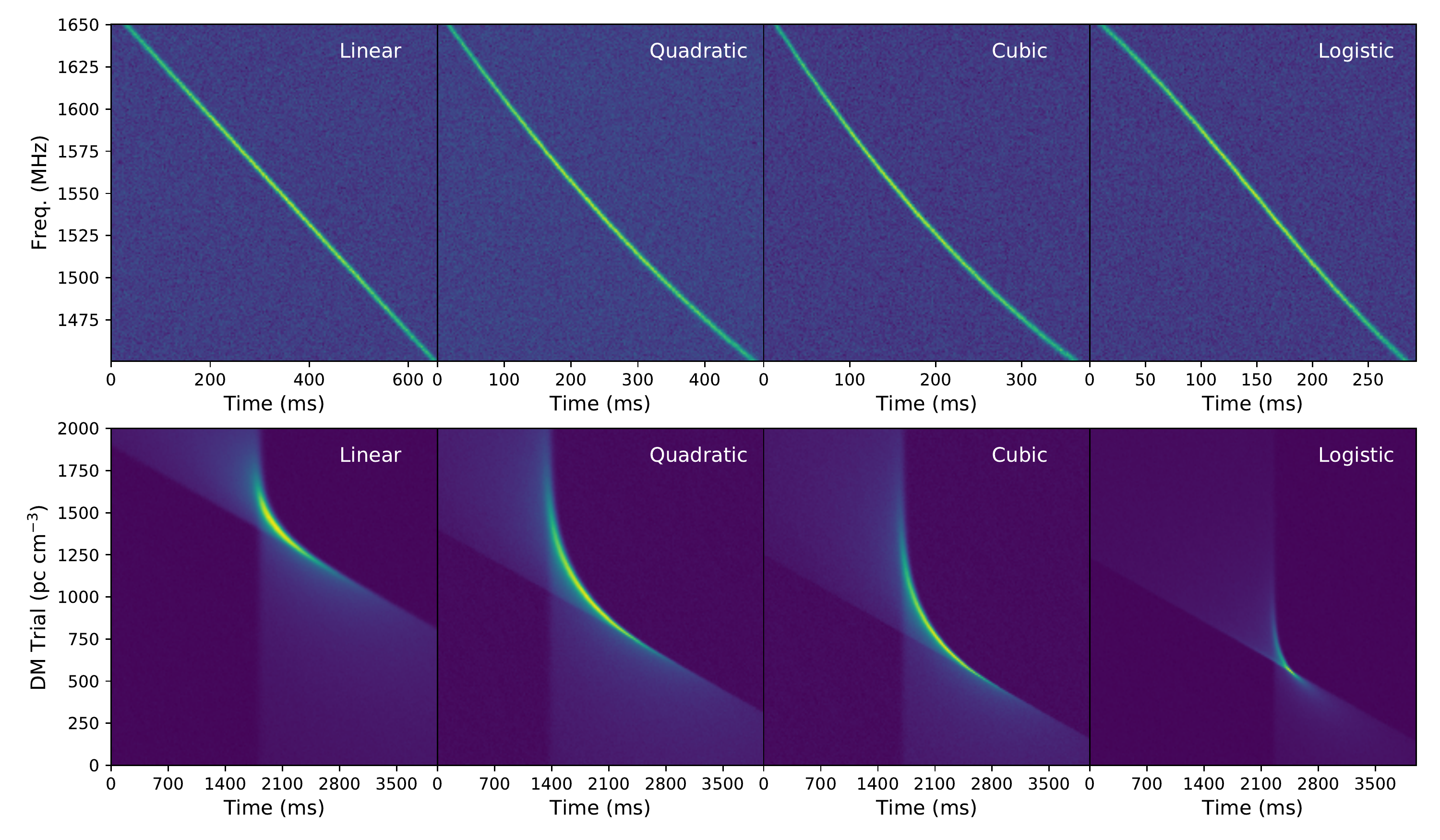}
    \caption{Top row: examples of simulated pulses with different dispersion
    relations: linear, quadratic, cubic, logistic (left to right). Bottom row:
    DM-space plots of the simulated pulses for a receiver model of $\Delta
    \nu_{\textrm{obs}} = 200$~MHz centred at $\nu_{\textrm{obs,c}} = 1550$~MHz.
    Initial starting time is arbitrary and different between plots.
    }
    \label{fig:sim_pulse_spectra_dm}
\end{figure*}

We model this effect by simulating a number of dispersion and receiver models at
L-band.  Pulses are modelled with linear ($\nu^{-1}$), quadratic ($\nu^{-2}$),
cubic ($\nu^{-3}$), and logistic ($(1+e^{-\nu})^{-1}$) functions which
cover a bandwidth from $\nu_{\textrm{p,0}} = 1300$~MHz to $\nu_{\textrm{p,1}} =
1800$~MHz.  The quadratic dispersion model is used as the reference, while the other
dispersion models are simulated as they are similar to anthropogenic signals of
known and unknown origin.  Example pulses are shown in the top row of Figure
\ref{fig:sim_pulse_spectra_dm}. The time duration of the pulse $\Delta
t_{\textrm{pulse}}$ is fixed such that the quadratically dispersed pulse has an
\gls{snr}-maximized DM of approximately 1000~pc~cm$^{-3}$. The pulse is
convolved with a Gaussian function to give a pulse width $\Delta
t_{\textrm{width}}$ of 3~ms. The amplitude of the pulse is set to unity across
the extent of the band.

\begin{figure}
    \includegraphics[width=1.0\linewidth]{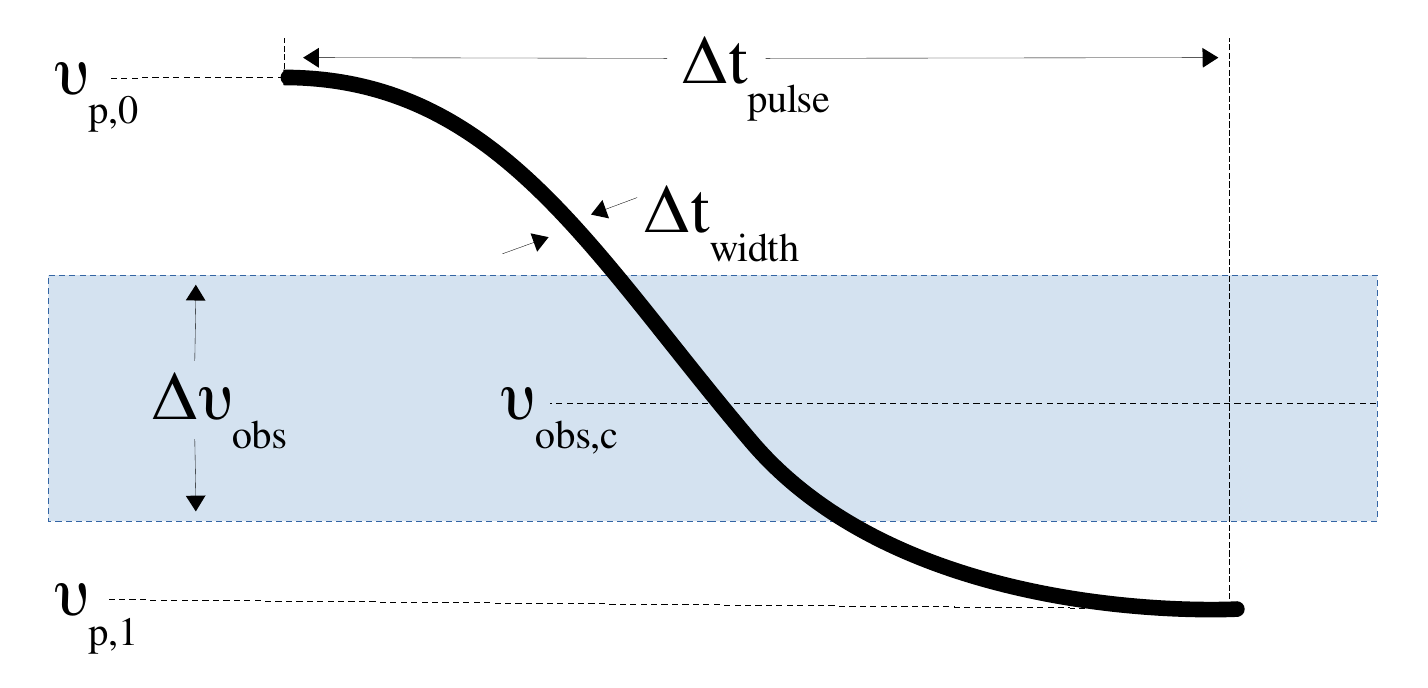}
    \caption{Logistic function pulse and receiver model used to simulate the
    response and \gls{snr}-maximized DM search. The pulse is modelled as a pulse
    of width $\Delta t_{\textrm{width}}$ that extends over a time period $\Delta
    t_{\textrm{pulse}}$, with a start $\nu_{p,0}$ and stop $\nu_{p,1}$ frequency
    beyond the bandwidth $\Delta \nu_{\textrm{obs}}$ of the receiver system
    centred at $\nu_{\textrm{obs,c}}$. An similar model was used for the
    linearly, quadratically, and cubically dispersed pulse models.
    }
    \label{fig:simulation_diagram}
\end{figure}

A receiver model is used to simulate the observation. This model is
parameterized by a central observing frequency $\nu_{\textrm{obs,c}}$, bandwidth
$\Delta \nu_{\textrm{obs}}$, frequency resolution $\Delta \nu_{\rm chan}$, time
resolution $\tau_{\rm int}$, and per channel noise $\sigma_{\textrm{chan}}$. The
bandpass response is modelled as a Gaussian with width $\Delta
\nu_{\textrm{obs}}$. A schematic of a simulated pulse and receiver model is
shown in Figure \ref{fig:simulation_diagram}. For the simulation, we sampled
centre observing frequencies from 1400~MHz to 1700~MHz in 20~MHz steps and
receiver bandwidths of 50, 100, and 200~MHz. The quadratically dedispersed
pulses ranged in \gls{snr} from approximately 20--40 depending on $\Delta
\nu_{\textrm{obs}}$. The time and frequency resolution of the data was
$\tau_{\rm int} = 256\,\mu \textrm{s}$ and $\Delta \nu_{\rm chan} = 360$~kHz.

\begin{figure}
    \includegraphics[width=1.0\linewidth]{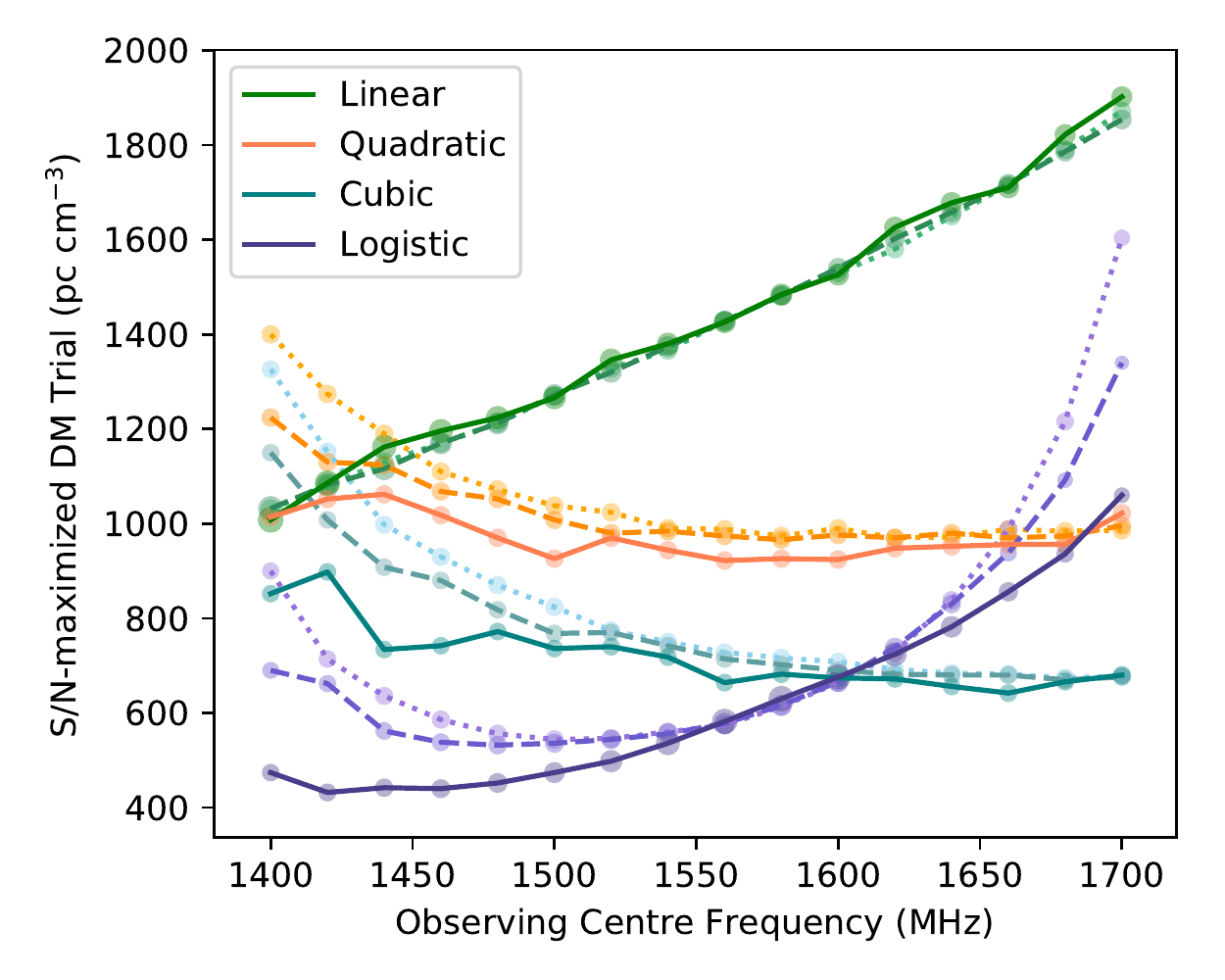}
    \caption{\gls{snr}-maximized DM trial for the simulation of pulses centred
    at 1550~MHz over a range of central observing frequencies. Pulses with
    linear (green), quadratic (orange), cubic (teal), logistic (purple)
    dispersion relations are simulated using a detection receiver with a
    bandwidth of 50~MHz (dotted), 100~MHz (dashed), 200~MHz (solid). The point
    size scales are the detection \gls{snr} relative to the quadratic pulse
    detection \gls{snr}.
    }
    \label{fig:sim_snr_max_dm}
\end{figure}

A \gls{dm} search of these simulated pulses results in significant detections in
compact regions of the DM trial space (bottom row of Figure
\ref{fig:sim_pulse_spectra_dm}). Each pulse was detected with similar \gls{snr}
to that of the quadratically dispersed pulse using the same simulation
parameters, but at different \glspl{dm} (Figure \ref{fig:sim_snr_max_dm}). The
\gls{snr}-maximized \gls{dm} of quadratically dispersed pulse stays fixed across
the choice of observing frequency as expected. Though, smaller observing
bandwidths result in an increase in the \gls{snr}-maximized \gls{dm} at lower
frequencies due to the pulse width. The cubically dispersed pulse shows a
similar flat response as the quadratically dispersed pulse but with a faster
upturn in \gls{dm} at lower $\nu_{\textrm{obs,c}}$.  The \gls{snr}-maximized
\gls{dm} of the linearly dispersed pulse has a linear response as a function
observing frequency, independent of observing bandwidth.  The
\gls{snr}-maximized \gls{dm} of the logistically dispersed pulse varies
significantly as a function of observing bandwidth and frequency. This is only a
small set of simulated parameters and dispersion models, but a wide range of
\gls{snr}-maximized \glspl{dm} can be produced which appear similar to the
quadratically dispersed pulse.

If these pulses are detected with a high \gls{snr} across a broad bandwidth then
a dispersion relation can be fit to the pulse to show if it follows a cold
plasma relation. But, in the case of a low \gls{snr} detection, or a
band-limited signal it is difficult to determine the dispersion relation. A
terrestrial pulse can be generated with a range of dispersion relations
resulting in a detection which can be difficult to differentiate from an
astrophysical source. Beyond using the test of a high \gls{snr} detection in an
\gls{frb} pipeline a number of additional tests can be done to verify a source
as astrophysical.

\section{FRB Verification}
\label{sec:verify_crit}

\begin{figure}
    \includegraphics[width=1.0\linewidth]{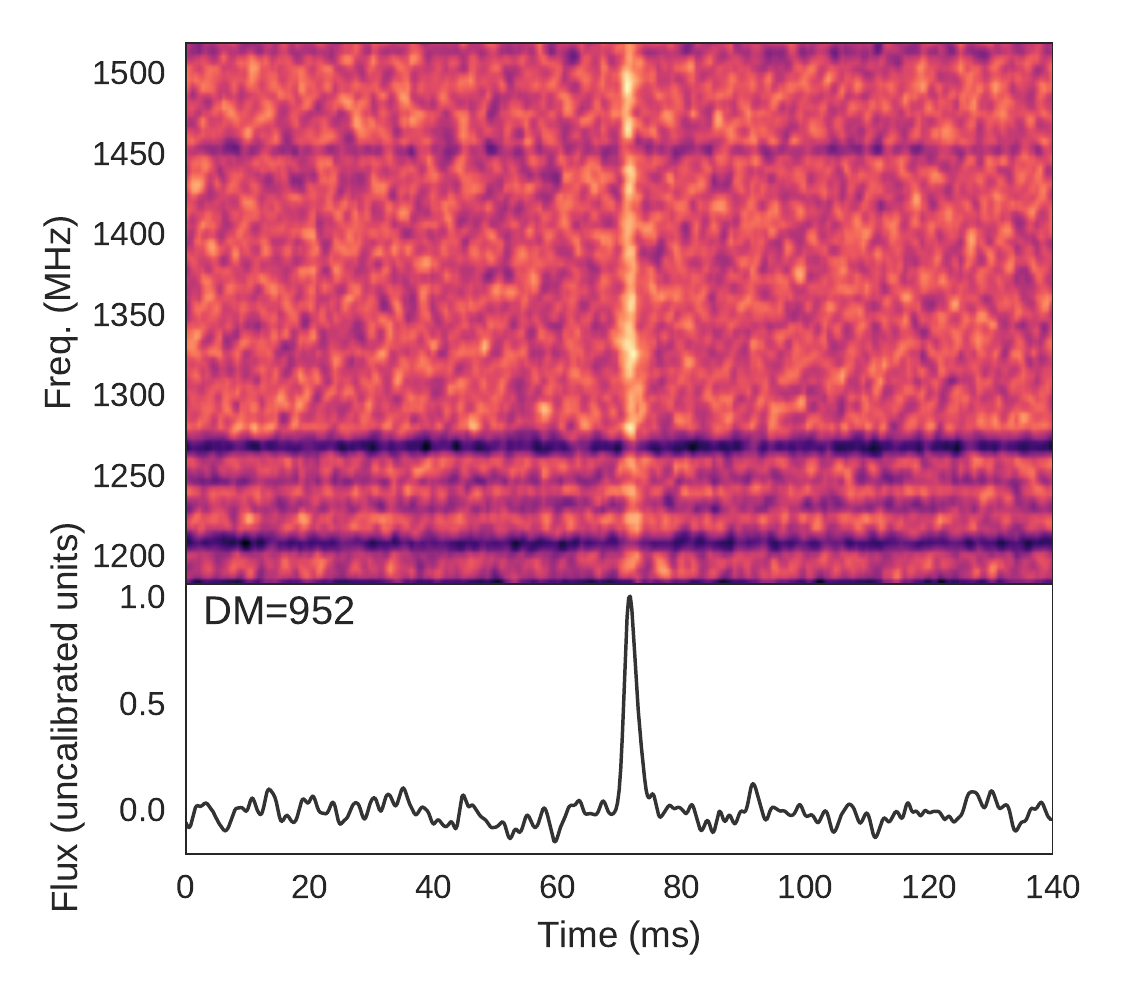}
    \caption{FRB\,130626 is a prototypical FRB with a broad-in-frequency,
    narrow-in-time, single component pulse at a DM of 952~pc~cm$^{-3}$ which is
    well in excess of the Galactic DM contribution in that line of sight.
    }
    \label{fig:FRB130626}
\end{figure}

Our ability to discover \glspl{frb} depends on how well we can describe what an
FRB is, and on having the data available to search for and verify events. There
is currently no formal definition, rather a set of community practices. These
vary between groups and instruments, but usually include information drawn from
images of power in the time-frequency plane and the time-DM plane, information
about the telescope state, information related to the presence of the signal in
one or more beams.  These standards are set in order to compare each new
detection to an \gls{frb} prototype: a broad-band signal, excessively dispersed
compared to a Galactic source along this line of sight, following a $\nu^{-2}$
dispersion relation, narrow in pulse width and possibly scattered following an
approximately $\nu^{-4}$ relation, such as FRB\,130626, as seen in Figure
\ref{fig:FRB130626} and \citet{2016MNRAS.460L..30C}. To this set of
characteristics it is important to add polarization, which is measured now in
most surveys. Although a clear picture is yet to emerge regarding the
prototypical polarization properties of \glspl{frb}, polarization can certainly
help indicate non-astrophysical origin of certain signals.  We do not consider
repetition a prototypical characteristic.  Though FRB~121102 is the only
verified astrophysical \gls{frb} because it repeats, all other detections are
apparently one-off events.  Further detections and follow-ups would help to
refine this characteristic.

This process of comparison follows the prototype theory of categorization
\citep{ROSCH1976382} which suggests that certain examples of a category are more
prototypical than others. A category will contain examples which deviate from
the prototypical as the prototype is an exaggeration of features deemed
characteristic to the category. We have defined a simple FRB prototype based on
available category evidence. Further detections will lead to refinement of this
prototype, and possible addition of new categories rooted in physical models.    


\subsection{Data}
\label{sec:detect_report}

The one-off nature of \glspl{frb} makes it essential to do sufficient
due-diligence when reporting on a detection or triggering a follow-up
observation to verify a true-positive detection. Over the past decade of
\gls{frb} surveys, a number of techniques have been developed to efficiently
filter the total power, time-frequency data to look for dispersed pulses.
Additional and important tests require capturing more data. It is not always
possible to capture sufficient data for every desired test, nor is it possible
with every observational setup.  The effort is, however, justified, as
questioning a detection from many possible angles will lead to higher confidence
in reporting an event as astrophysical or rejecting it as not interesting. 

If most \glspl{frb} are indeed one-off events, then the detection data are the
only data that will be available. Publication of the detection data allow
independent verification, and can be used as input to test independent
pipelines. The following should at least be available:

\begin{enumerate}
    \item A dynamic spectrum which fully encompasses the extent of the dispersed
    pulse. Preferably, both the raw dynamic spectrum and the \gls{rfi} flagged
    and/or normalized spectrum computed in the detection pipeline.
    \item Technical data such as time and frequency information, telescope
    pointing, and other telescope observing parameters.
    \item For a multi-beam system, dynamic spectra of each beam covering the
    extent of the pulse. Similarly for a \gls{tab} or interferometric detection.
    \item The detection parameters such as peak \gls{snr}, best DM, pulse width,
    and others.
    \item A list (or guide) of software (including the versions) and parameters
    used to generate diagnostic plots.
\end{enumerate}

These requirements are typically met in previously reported \gls{frb}
detections. Often the public data have been normalized and re-quantized to 2 bits
for long-term archiving. This reduction in the dynamic range results in
potential distortion of pulses, and normalization can hide instrumentational
effects. Where possible, the data at the full bit depth, time, and frequency
resolution should be stored.  The last point, a guide on which software was used
is often not reported.  Differences in software and data formats can produce
different results, such as the reported \gls{snr} or time definition. For most
of the Parkes detections, filterbank data are publicly available and FRBCAT
\citep{2016PASA...33...45P} provides a well-curated repository for information
and links to data sources.

To allow for a more complete set of criteria tests and to further improve the
evidence of an astrophysical detection, and in addition to the list above, we
propose capturing the following:

\begin{enumerate}
    \item Dynamic spectra which not only encompass the detected pulse, but
    include data over a longer time spans, e.g. a few minutes before and after
    the detection. Similarly for multiple feeds, additional \glspl{tab}, and
    interferometric baselines.
    \item Raw voltage or complex spectral data before power detection and
    integration that retains time and frequency resolution, and phase
    information.
\end{enumerate}

A dynamic spectrum extended in time, can be used to find gain and band pass
variations. Low-\gls{snr} pulse searches can be performed to test if the source
repeats or if there is an anomalously high number of events at different DMs.  A
test observation of a known pulsar nearby in position and time to the detection,
perhaps in the beam during the detection, is a useful method to provide a
verification of the system state during a detection.

Complex voltages, though costly to capture and store, provide valuable insight.
They allow for coherent dedispersion which reduces smearing, and increases the
detection sensitivity as in FRB\,180301 \citep{atel11376} where an apparently
narrow-band detection was revealed to have low-level flux across a larger band.
The time and frequency resolution can be adjusted to look for structure. If
multiple sites or elements are used, the signals can be correlated for
coincidence detection and localization.

\subsection{Criteria}
\label{sec:criteria}

For verification of new FRBs, we propose a list of criteria of two types: the
first type tests the similarity between new detections and the prototype, and
the second type is used to preclude terrestrial origins of the signal. Together,
these two sets provide a framework, which should be used to state the confidence
of any new \gls{frb} discovery. This list of criteria will improve with time; as
new events are discovered, both astrophysical and terrestrial, new tests can be
included. The criteria we propose here have been primarily developed from single
dish surveys.  Use of interferometric arrays will provide powerful further
criteria, particularly relevant to localization of signals in the sky.

Included in the criteria are tests to verify whether the telescope was
functioning as expected. This is to say that there is a prototypical status for
each telescope and receiver system, parametrized by observable quantities, which
the data obtained during a detection can be compared against.  This comparison
requires the availability of the necessary data and expert knowledge of the
observer and telescope operators. However, it should be done explicitly when
reporting a new detection.

During a search for anomalous signals we expect a number of reported
false-positives events, such as Perytons \citep{2011ApJ...727...18B} and the
events presented in Section \ref{sec:false-pos}.  Public datasets for events
that are confirmed as false positives is also extremely useful in defining
criteria for classification. 

\subsubsection{Criteria and scoring suggestions}

We define the two types of verification criteria stated above as `Similarity to
the prototype' and `Terrestrial in origin'. For each criterion, we propose 6
possible answers, which we colour-code in our own categorization below. Answers
1 to 4 are used for the responses: `Identical', `Similar', `Not similar',
`Completely different', while 5 and 6 are used for the responses `Data not
available' and `Test not performed' respectively.

In the following, we provide a set of questions to help assign a numerical
response of 1-6 for each criterion. We have also gone through this process
ourselves for a number of published FRBs and other spurious detections, thus
providing reference responses for each criterion.

Although it would be tempting to prescribe a combined score, based on which a
signal can be classified as an astrophysical FRB or not, the known population is
not numerous enough to correctly express all possible true characteristics of
the class. At this moment in time, the scores serve, as we demonstrate in the
following, to assess individual sources. As the population increases, following
this framework of scores should lead to more certain classification. In
particular, we have considered carefully the effect of ascribing high scores to
values of metrics that fall within the range of verified events. Our reasoning
for doing this is that, if a source is scored low for some particular criterion
in the following list, but scores highly overall, this only serves in expanding
the definition of the prototype. If a source scores low in most criteria, and
therefore also overall, it would be hard to objectively classify it as an FRB.

\paragraph{Signal to Noise}

What is the measured \gls{snr} of the dedispersed, and frequency-collapsed
pulse? Can this \gls{snr} be verified? The actual \gls{snr} could be different
from the reported \gls{snr} (see Section \ref{sec:low_snr}). The highest scores
should be given for a reliably measured \gls{snr} that is both significantly
high and within the expected range based on verified FRB detections.
Progressively lower scores should be given if the \gls{snr} is too low, not
verifiable, and/or significantly outside the expected range.

\paragraph{Boresight Flux}

What is the implied flux based on the detection \gls{snr} and telescope
sensitivity, assuming the detection occurred at boresight?  Is this within the
range of previously reported fluxes? How does it compare to the statistical FRB
flux distribution?  The majority of reported \glspl{frb} have an implied
boresight flux between $0.2-2$~Jy, although there is a tail of bright
\glspl{frb} potentially exceeding 128~Jy \citep{2016Sci...354.1249R}. Extremely
bright \glspl{frb} (>~kJy) are not expected based on the extensive previous
non-detection results from low sensitivity, large sky coverage surveys
\citep{2012ApJ...744..109S,2015MNRAS.452.1254K,2016MNRAS.458.3506R}. However,
extremely bright \glspl{frb} can not be completely discounted. The highest score
should be given to FRB candidates with fluxes within the range of known sources,
and the lowest scores to FRB candidates with the greatest deviation from the
known and verified population.

\paragraph{Pulse Width}

What is the pulse width of the dedispersed and frequency-collapsed pulse? Is it
within the range of previously reported widths? Reported widths range from tens
and hundreds of microseconds (FRB\,150807, FRB\,170827) up to tens of
milliseconds (FRB\,170922, FRB\,160317). The lower pulse width limit is based on
the time resolution of the instrumentation, and the pulses may indeed be
unresolved.  There may be astrophysical \glspl{frb} with wider pulse widths, but
most historical surveys have had a maximum time-binning scale on the order of
tens of milliseconds. FRB\,121102 shows a wide range of pulse shapes and widths
\citep{2018Natur.553..182M,atel10675}. The highest scores should be given for
signals with widths within the range of verified FRB detections.

\paragraph{High-resolution Structure}

Search pipelines identify candidates by applying a filter to the dedispersed
data. This hides any potential high time and frequency structure that may be
present, as in the case of radar signals (Section \ref{sec:LOFAR_RADAR}). The
dynamic spectrum should be checked for sub-structure at both the highest time
and frequency resolution. Complex voltages allow for coherent dedispersion and
flexibility in channelization and integration, which provides further
sensitivity in searching for structure. The highest scores should be given to
signals whose high-resolution structure can be explained by a natural process
and resembles the structure of verified FRBs. Progressively lower scores should
be given as the ``distance'' from the norm increases.

\paragraph{Multiple components}

Are there multiple components to the pulse? And how many? Does this number
change when dedispersing with different DMs? Most \glspl{frb} appear as single
component pulses. But, a few, such as FRB\,130729 and some bursts from
FRB\,121102, show multiple components, which can lead to differences in the
`optimal' DM \citep{2018Natur.553..182M}. Pulses from FRB\,121102 often contain
multiple components. As such, it deviates from the prototype. As FRB\,121102 is
known to be astrophysical, this criterion can not invalidate a detection. If the
number of components resembles verified FRBs and other known astrophysical
pulses, a high score should be given.

\paragraph{Broad-band}

Does the pulse extend in frequency to the edges of the observing band, bottom
edge, top edge, both, or neither? A pulse that extends across the full observing
band fits the prototypical \gls{frb} model. There are now examples of
narrow-band (within the observing band) FRBs (such as FRB\,121102), as well as
narrow-band terrestrial signals, such as radar (Section \ref{sec:LOFAR_RADAR}),
so not having a broad-band signature does not entirely rule out an FRB.  We
expect this criterion to be improved by future telescope back-ends offering
broader bandwidths. Higher scores should be attributed to signals that cover
greater extents of the band, in a way that can be explained by a physical
process and resembles other verified signals.

\paragraph{Spectral Index}

Is there a measurable apparent spectral slope? Deconvolving the instrumental
response from the astrophysical spectral response is difficult.  Depending on
the (mostly unknown) position of a detected pulse within the telescope beam,
there will be a frequency-dependent sensitivity response, which will induce
chromaticity.  If the pulse appears band-limited with a steep spectral
index, this could indicate the pulse flux drops below the system noise in parts
of the band. If a shallow spectral index is fit to a band-limited pulse, this
could indicate the pulse is intrinsically band-limited. Higher scores should be
attributed to signals whose spectral slope falls within the range of possible
spectral slopes of verified FRBs. Exceedingly steep, in both sense, spectra,
should be attributed lower scores.

\paragraph{Scattering}

Does the pulse appear to be scattered and/or scintillating?  Accurate scattering
and scintillation measurements are evidence for radio waves having passed
through an inhomogeneous \gls{ism}, and can help infer properties of either the
host galaxy of the FRB,  the media surrounding it, or the \gls{igm}.

Scattering timescales are estimated, as in the case of FRB\,130626
\citep{2016MNRAS.460L..30C}, by fitting an exponential tail ($\propto
e^{-t/\tau}$, typical of a thin scattering screen) to the scatter-broadened
pulse. The characteristic timescale, $\tau$, is expected to have a strong
frequency dependence ($\tau \propto \nu^{-\alpha}$), with most models predicting
a power law frequency scaling index $\alpha =4$ or $4.4$ (e.g.
\citealt{Rickett1977}), and observations of pulsars often showing $\alpha < 4$
(e.g. \citealt{Lewandowski2015,Geyer2017}).  Spectral indices computed from
broad band FRBs, should be compared to these expected values.  Measurement of a
scattering tail with a spectral index in the range mentioned above warrants the
designation of `Identical', while gradual departure from this picture should be
described using the other responses. 

Comparing obtained $\tau$ values to values predicted by current
electron density models for the Milky Way (e.g.
\citealt{2002astro.ph..7156C,2017ApJ...835...29Y}) also provides a
test for the extragalactic nature of a detected signal. The closer the
scattering characteristics are to what is seen in the verified FRB
population, the higher the score should be for this category.

\paragraph{Scintillation}

As with scattering, scintillation may also occur due to passage through an
inhomogeneous medium. Scintillation in pulsars manifests itself as an organized
pattern of time and frequency modulation of the total intensity. For FRBs, the
scintillation timescale may be longer than the duration of the pulse, hence not
sampled. The scintillation bandwidth on the other hand may be observable due to
the broad band systems used for searching.  Scintillation bandwidth ($\Delta
\nu_d$) estimates are obtained by computing the FWHM of the auto-covariance
function of the spectrum.  The $\Delta \nu_d$ measurement corresponds to a
broadening (scattering) timescale of $\sim 1/(2\pi\Delta \nu_d)$.  

The highest score should be attributed to sources whose scintillation bandwidth
is consistent with a measurement of scatter-broadening, and lower scores for
lower consistency. A non-detection of either scattering or scintillation is
considered `Similar' (i.e. neutral), until an improved set of categories
emerges. 

As seen in the XAO event (Section \ref{sec:xao_event}),
frequency modulation can appear similar to scintillation, hence this criterion
should be used with care and in conjunction with scattering measurements. 

\paragraph{Polarization Characteristics}

Were full Stokes parameters measured for the pulse? Does the pulse show
polarization characteristics? Can a rotation measure be fit to the pulse?
Positive answers to these questions should result in a high score.  Multiple
\glspl{frb} have been reported to be polarized
\citep{2015MNRAS.447..246P,2015Natur.528..523M,2016Natur.530..453K,2016Sci...354.1249R,2017MNRAS.469.4465P,2018MNRAS.478.2046C}.
For example, FRB\,121102, appears to be linearly polarized after the extreme
\gls{rm} is corrected for \citep{2018Natur.553..182M}, and FRB\,110523 is
measured to have a high fraction ($\sim$ 44\%) of linear polarization. Where
possible, it is insightful to compare RM values to RM measurements of nearby
pulsars. The RM estimate for FRB\,110523 was compared to an observation of PSR
B2319+60, within 2 degrees on the sky, using the same pipeline.  \gls{rfi} has
complex polarization characteristics, such as Phase-shift keying
(\cite{horowitz2015art}, Chapter 13), as a way of encoding information.  These
encoding schemes will appear different from polarized astrophysical sources, and
will not be well fit by a Faraday rotation model. Low scores should be
attributed to candidates with artificial polarization characteristics, and
higher scores to candidates where the polarization can be understood by the
aforementioned physical processes.

\paragraph{DM Excess}

What is the \gls{snr}-maximized \gls{dm}? How does this compare to the modeled
Galactic DM along the line of sight? Is it well in excess of the Galactic
contribution? This is a standard test to differentiate between a Galactic and
extragalactic source. If the ratio of measured \gls{dm} to Galactic model
\gls{dm} is at least a factor of two then the source is likely extragalactic.
Otherwise the source is likely a pulsar or \gls{rrat} within the Galaxy.  If
there are multiple components, what is the `component-optimized' DM, how does it
differ from the \gls{snr}-maximized DM? When there are multiple components to a
pulse, the choice of DM can result in different frequency-averaged pulses such
as in FRB\,130729 in which the choice of DM can result in a single component or
multi-component pulse profile (see Section \ref{sec:frb_signals}). Lower scores
should be attributed to candidates whose \gls{dm} could be explained as
Galactic.

\subsubsection{Terrestrial Origins}

\paragraph{Dispersion Relation}

Does the pulse follow a cold plasma $\nu^{-2}$ dispersion relation? If there is
sufficient \gls{snr} in individual time-frequency bins across a significant
fractional bandwidth of the dynamic spectrum then a dispersion relation fit
showing a $\nu^{-2}$ relation is evidence for a dispersed, natural source.
Conversely, a fit that diverges from $\nu^{-2}$ is good evidence for an
artificial source (Section \ref{sec:xao_event}). Performing model selection
between a $\nu^{-2}$ relation model and other models can be done to
statistically compare models.

\paragraph{DM Trial Space}

In a DM trial space search (positive and negative DM trials) are there other
high-\gls{snr} events nearby in time to the detected pulse? If there are \sout{few}
other events \sout{above} during the time window then that is evidence for a true
detection. But, if there are events seen at similar \gls{snr}, especially in the
negative DM space, then it could be that the \gls{rfi} environment has
increased the false-positive event rate.

\paragraph{Repeating Events}

Repeating events are evidence for either an astrophysical or a terrestrial
origin. Most \glspl{frb} appear to be one-off events, but if the source does
repeat then it can be localized or verified with another telescope. If the
source repeats at different points such as the \gls{xao} event, then the source
is terrestrial.

Were follow-up observations performed to search for repeating events?  Was a
lower-\gls{snr} search performed around the detected \gls{dm} in the survey
data? If the source does repeat, what was the time scale? Could a periodicity
search be performed? And, did the repeating events occur at different telescope
pointings?

The current prototype is considered to be a once-off burst. We label the
repeating FRB~121102 as `Not Similar' to the prototype. Further detection of
repeating FRBs will facilitate changing the prototype model. 

\paragraph{RFI Environment}

An excess in \gls{rfi} during a detection compared to the typical \gls{rfi}
environment reduces the confidence in an astrophysical detection.  Was the
\gls{rfi} environment drastically different during the event detection compared
to typical observations? What was the effective bandwidth of the receiver after
\gls{rfi} flagging? If the \gls{rfi} environment was different, there could be
an increase in the number of false-positive events, likely seen in the DM trial
space test. Or, the reduced bandwidth can limit a $\nu^{-2}$ dispersion relation
fit.

\paragraph{Telescope State}

While on the surface a trivial criterion, it is important to check that the
telescope was in a valid state. For example, was the analog signal chain in a
linear state, was the feed in the correct position, was other equipment active
that could cause local interference? This is a site and telescope-dependent
criterion, which requires expert knowledge of the system in order to test this
criterion with due diligence.  A regular test of the overall telescope state
would be to observe a known pulsar during observations.

\paragraph{Bandpass Variation}

During the event, does the bandpass shape appear similar to the expected
bandpass seen in typical observations? Most search pipelines apply a bandpass
correction to normalize the noise variance. This flattens the bandpass, and
hides any changes in the measured bandpass response. The first indication that
the electronics were not functioning correctly in the ALFABURST event (Section
\ref{sec:D20161204}) was the change in bandpass response.

\paragraph{Gain Stability}

Does the gain show similar variance over a window of time around the event
compared to the gain variation seen in typical observations? Similar to bandpass
normalization, a high-pass filter is often applied in a search pipeline to
reduce long-term gain variations. This may be hiding differences in gain
variation during an event detection. The gain stability of a system during an
event detection should be similar to the stability on a longer time scale around
the detection.

\paragraph{Telescope Pointing}

Where was the telescope pointing in the local reference frame? Was it near the
horizon or known \gls{rfi} sources? Weak \gls{rfi} sources on the horizon are
picked up by the high forward gain of dish telescopes. And, strong \gls{rfi}
sources can still be picked up in the side and back-lobes of the beam, or
directly illuminate the feed.

Were there known satellites in the primary or side-lobes of the beam? For
example, geosynchronous satellites are located within $\pm 15^{\circ}$
declination \citep{anderson15operational}.  Publicly available satellite orbital
parameters can be used to determine the presence of a satellite near a telescope
pointing position.  The transponder frequencies and positions of commercial
satellites are regularly maintained, though not all are publicly reported, e.g.
military satellites. The presence of a satellite operating at frequencies in the
observing band will result in an increase in the overall system gain. But, if
the gain has been normalized the source might not be apparent.

\paragraph{Local Time}

What was the local time of the detection? Was there regularly scheduled
maintenance? Does the time coincide with an increase in local \gls{rfi}, as in the
case of Perytons (Section \ref{sec:perytons})?  Or, for a commensal system, what
was the primary observing schedule?  It could be that equipment was active which
typically is not during standard observations, such as the \gls{alfa} candidate
(Section \ref{sec:D20161204}).

\paragraph{Multi-beam}

Is the receiver a multi-beam system? Was the pulse detected in multiple beams?
If so, is there a difference in \gls{snr} between beams?  Multi-beam systems, in
particular the Parkes multi-beam receiver, have been very successful in the
detection of \glspl{frb} as they provide good evidence for an event occurring in
the far-field versus in the near-field.  A multi-beam system can be used to
perform coincidence detection to significantly reduce the number of
false-positive detections.  FRB~010724 \citep{2007Sci...318..777L} was detected
in multiple beams of the Parkes multi-beam receiver, but at different
sensitivities indicating that the source, though extremely bright, was localized
to a position in the sky.  Perytons were also detected with the Parkes
multi-beam receiver but in all beams at similar sensitivities indicating the
source was local \citep{2011ApJ...727...18B}.

\paragraph{Tied-array Beam (\gls{tab})}

Was the pulse detected with a \gls{tab} using an array of elements? If so, was
the pulse seen in individual elements? Were multiple \glspl{tab} active during
the observation? If so, was the pulse seen in multiple beams?  Similar to a
multi-beam detection, a tied-array beam detection can be used to determine if
the source was in the near-field. Further, the near-field limit of a tied array
is typically much larger than that of a dish. For example, the near-field limit
of Parkes is $\sim 40$~km, while \cite{2018MNRAS.478.1209F} set a near-field
limit of the source of FRB~170827 to $\geq 10^4$~km with a tied-array detection
which can rule out the source being local up to medium Earth orbits.  If the
individual element data are available, then subsets of elements can be combined
to show the source was detected in most elements rather than just in one or a
few.

\paragraph{Interferometric Array}

Was the pulse detected while baseline correlations were recorded in an
interferometric mode? Was the pulse localized within the primary beam? Is the
pulse detected on individual baselines?  Similar to the tied-array criterion, an
interferometric detection can determine if the source was in the near field. An
\gls{frb} detection on baselines $\geq 10$~km at L-band frequencies would be
sufficient to determine if the source is a terrestrial satellite.
Interferometric detection of a pulse provides strong evidence by removing
element auto-correlation effects. Though, local \gls{rfi} will persist,
synthesis imaging can be used to localize the source.

\paragraph{Multi-site Observations}

Was the pulse detected in a multi-site observation campaign? If so, was the
pulse detected at multiple sites? Multi-site detection is very strong evidence
for a pulse being astrophysical.

\section{Application of FRB criteria to previous detections}
\label{sec:appln_to_previous_detections}

\begin{figure*}
    \includegraphics[width=1.0\linewidth]{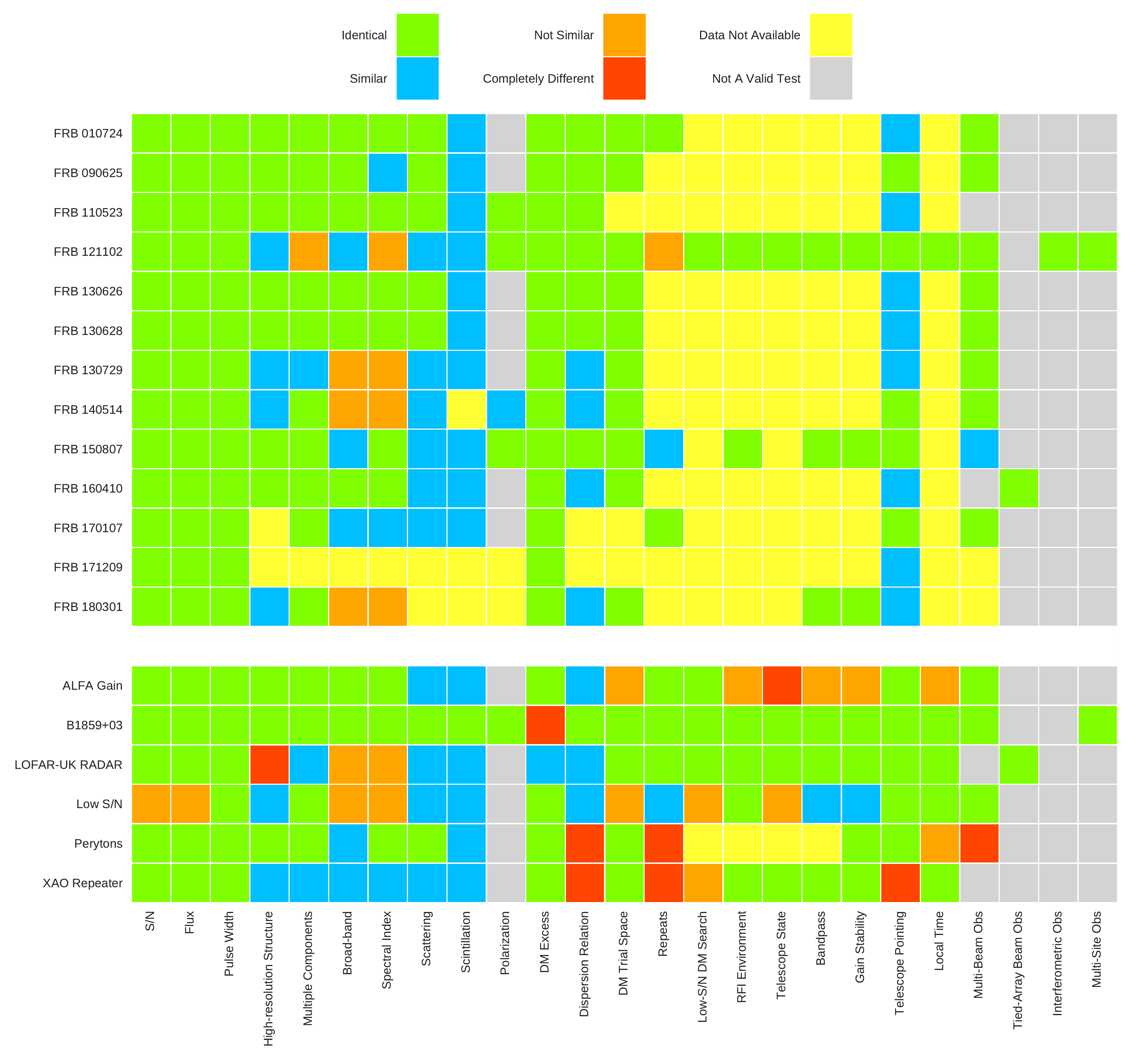}
    \caption{Verification criteria (Section \ref{sec:criteria}) heat map of some
    of the previously reported FRBs (top) and the terrestrial sources discussed in
    this work (bottom).  Green indicates a criterion test is identical to the
    prototypical detection or observation, thus providing evidence in favour of
    an astrophysical origin.  Blue indicates a criterion test result is similar
    to the prototype, but is not identical, and thus, is neutral evidence of an
    astrophysical origin.  Orange indicates the criterion results deviates from
    the expected prototype result and the event may be of terrestrial origin.
    Red indicates the criterion result is completely different from the
    prototype, and the event is terrestrial in origin. Yellow indicates the
    criterion result could not be determined from the available data.  Grey
    indicates the criteria is not valid based on the observation.
    }
    \label{fig:heat_map}
\end{figure*}

\subsection{FRB signals}
\label{sec:frb_signals}

Having provided the FRB verification criteria above, we have re-analyzed the
previous reported \glspl{frb} based on the available literature and public data.
This analysis provides examples of answers 1-6 from Section \ref{sec:criteria}
for a number of sources. Figure \ref{fig:heat_map} captures the results in a
colour-coded table.  Overall, it is clear that the detections reported in the
literature score well against the prescribed criteria, which is to be expected
as many of these criteria are implicit to the observational definition of an
\gls{frb} such as high \gls{snr} and a line of sight dispersion measure in
excess of the Galactic contribution.

\begin{figure}
    \includegraphics[width=1.0\linewidth]{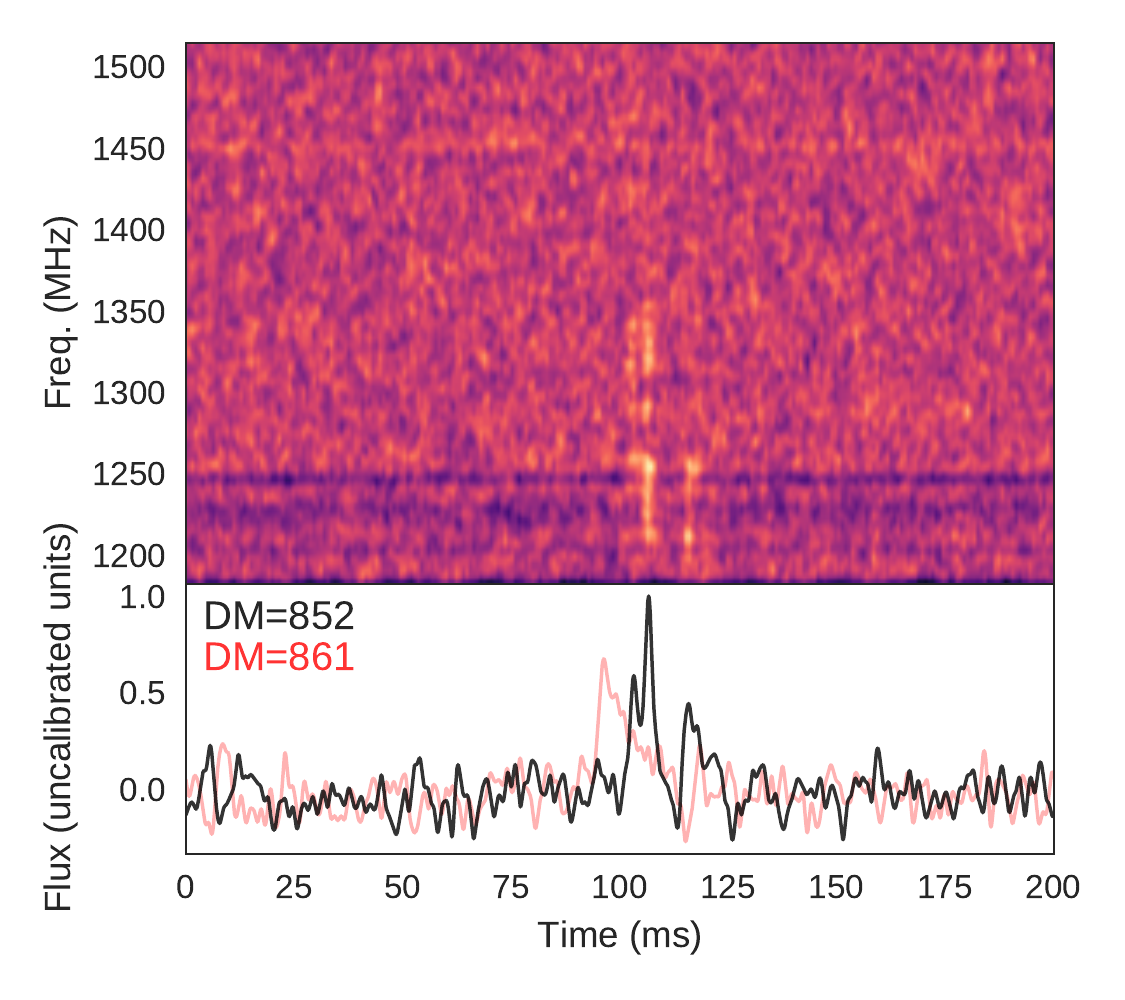}
    \caption{FRB\,130729 dedispersed with a DM of 852~pc~cm$^{-3}$ (black), this
    is different from the DM of 861~pc~cm$^{-3}$ (red) reported in
    \protect\cite{2016MNRAS.460L..30C}.  The detected FRB has two distinct
    components separated by approximately 10~ms. Data are presented at the native
    recorded resolution of 64~$\mu$s, 390~kHz convolved with a Gaussian
    smoothing filter of size 512~$\mu$s, 3.125~MHz.}
    \label{fig:FRB130729}
\end{figure}

Starting from the lowest scores in our sampled set, we note that FRB\,130729 is
band-limited to the lower half of the observing band, and there is a sharp
decrease in flux around 1350~MHz (Figure \ref{fig:FRB130729}). We are using a
dispersion measure of 852~pc~cm$^{-3}$ which is different from the originally
reported dispersion measure of 861~pc~cm$^{-3}$. Using this new dispersion
measure there are two distinct components. The original detection paper
\citep{2016MNRAS.460L..30C} notes that FRB\,130729 is potentially due to
terrestrial \gls{rfi}, which is made clear in the low criteria scores.

\begin{figure}
    \includegraphics[width=1.0\linewidth]{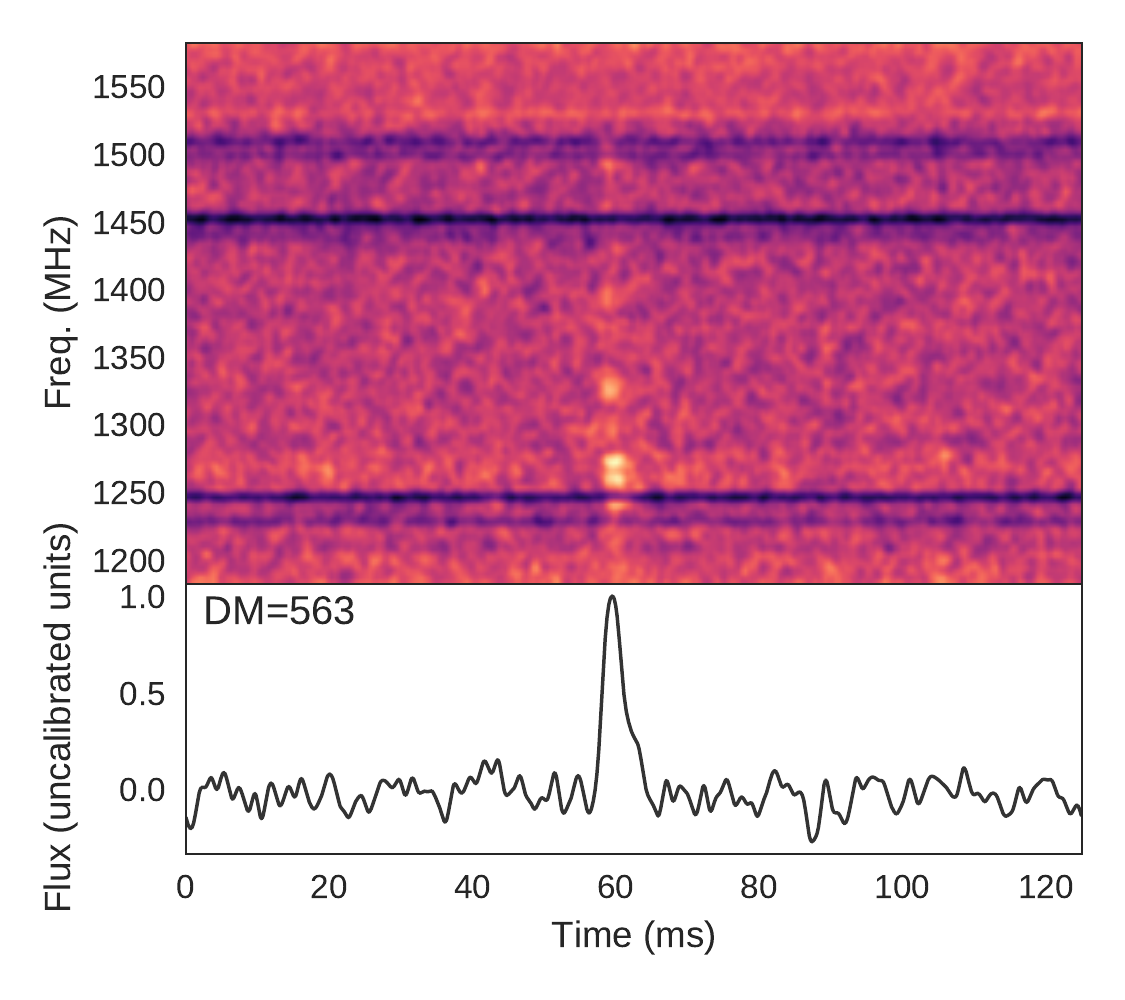}
    \caption{Dynamic spectrum of FRB\,140514 dedispersed with a DM of
    563~pc~cm$^{-3}$.  Data are presented at the native recorded resolution of
    64~$\mu$s, 390~kHz convolved with a Gaussian smoothing filter of size
    512~$\mu$s, 3.125~MHz.
    }
    \label{fig:FRB140514}
\end{figure}

\begin{figure}
    \includegraphics[width=1.0\linewidth]{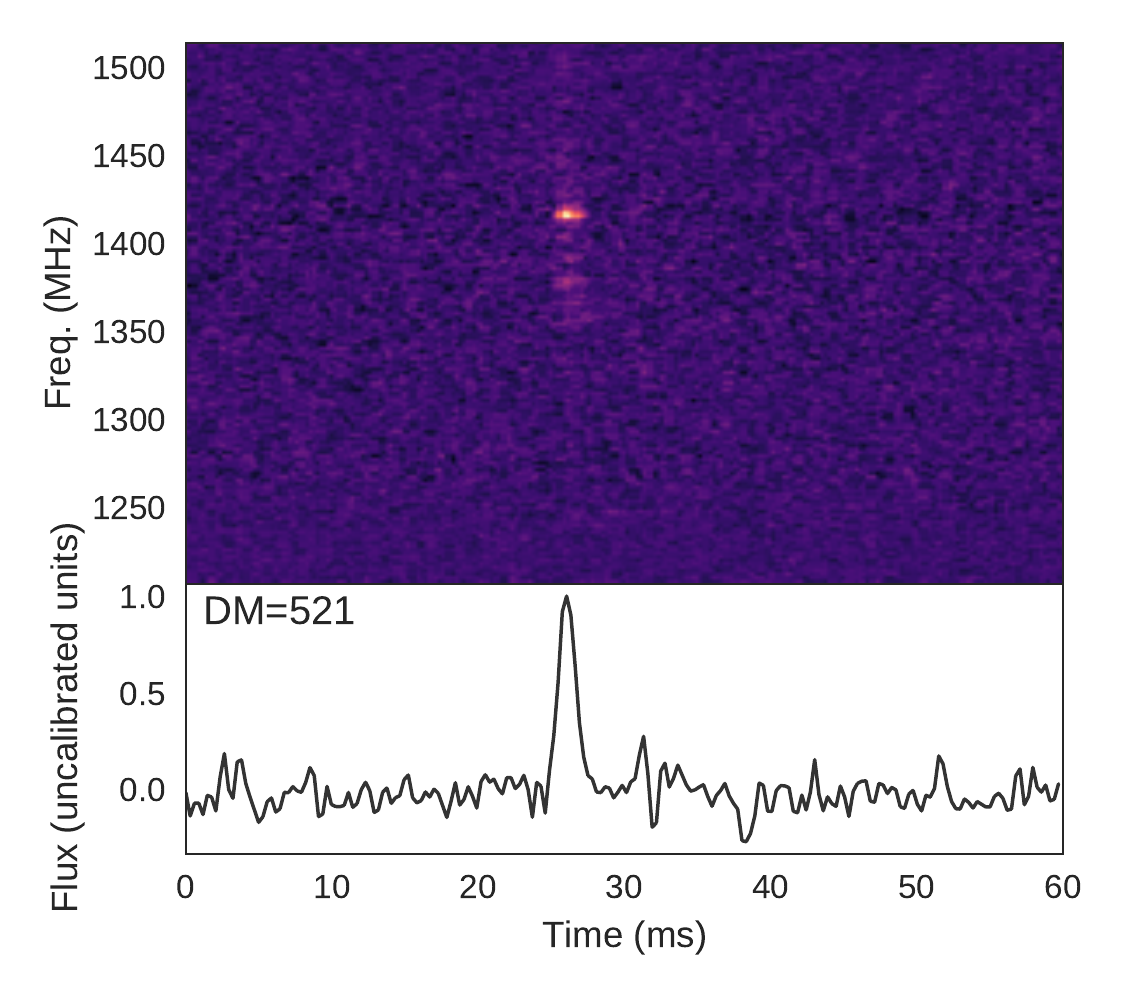}
    \caption{Dynamic spectrum of FRB\,180301 dedispersed with a DM of
    521~pc~cm$^{-3}$.  data are presented at the native recorded resolution of
    292.57~$\mu$s, 437.5~kHz convolved with a Gaussian smoothing filter of size
    29.2~$\mu$s, 875~kHz.
    }
    \label{fig:FRB180301}
\end{figure}

FRB\,140514 (Figure \ref{fig:FRB140514}) and FRB\,180301 (Figure
\ref{fig:FRB180301}) both score poorly in the broad-band criteria as the
majority of the detected flux is concentrated into narrow regions of the band.
This low score affects the dispersion relation criteria test as there is not
sufficient fractional bandwidth to perform a good fit. The frequency structure
of both of these detections could be due to scintillation or plasma lensing.
These low scores do not discount FRB\,140514 and FRB\,180301 as astrophysical, they
indicate that FRB\,140514 and FRB\,180301 diverge from the prototypical \gls{frb}
model.  We note that FRB\,121102 diverges significantly from the prototype model
as individual pulses vary in bandwidth, apparent scattering, spectral index, and
structure.  This is a useful example to show that the highest test scores are
not needed to verify a detection as astrophysical.

FRB\,010724 \citep{2007Sci...318..777L}, being the first reported detection, fits
the prototypical model well as many of the standard criteria we now use were
initially decided in the detection report.  Similarly, FRB\,110523, FRB\,130626, and
FRB\,130628 all match the prototypical model well, and thus show high scores in
most of the tests. FRB\,090625 shows variation in flux across the band, possibly
due to beam colorization or scintillation.  This is indicated by the lower
spectral index criteria score, but FRB\,090625 is otherwise also prototypical.
FRB\,150807 shows a drop in flux at the high end of the observing band due to a
steep apparent spectral index which is either intrinsic to the source or due to
the beam response. The reported detection included a number of system state
tests such as checking the gain and bandpass response, and the \gls{rfi}
environment which increases the overall score of this detection.

In most cases, detections do not report on the telescope state and environment
during the time of detection. Though data are made public in many of the
detections, these data have already been normalized or calibrated so it is not
possible to verify the telescope state. More recent detections are often lacking
in data or detailed analysis. For example, very little information on
FRB\,171209 has been reported, thus almost none of the criteria can be tested.
Providing public data not only allows for independent verification, but also for
new tests on previous detections. For example, the lack of public data of
FRB\,170107 means that criteria not reported in the initial detection can not be
tested.

\subsection{Terrestrial signals}

Also included in Figure \ref{fig:heat_map} are the results for the various
detections discussed in Section \ref{sec:false-pos}. We also include
PSR~B1859+03 as an example of single pulses from a pulsar during the ALFABURST
surveys.  In all these cases there is at least one failure.  The LOFAR radar
event shows artificial high-resolution structure.  Perytons and the XAO repeater
fail the dispersion relation test and repeat at different telescope pointings.
The ALFA gain variation has multiple negative test results relating to the
system response compared to the expected response.  The low \gls{snr} event does
not have a single critical test failure but fares poorly in several tests.  No
single test is sufficient to decide that an event is terrestrial. Figure
\ref{fig:heat_map} demonstrates how, given enough tests, a strong case can be
made to classify all these events as terrestrial.

\section{Future Observational Methods}
\label{sec:future_methods}

Beyond detection of more \glspl{frb}, the goals of current surveys are to
localize the source to host galaxies, and detect pulses across broader
bandwidths and different frequencies in the radio spectrum. Verifiability should
also be considered in these surveys. This means capturing additional data as
discussed in Section \ref{sec:detect_report} and additional observing methods.

The simultaneous detection of a signal with telescopes at multiple sites is
clear evidence for an astrophysical \gls{frb}.  Multi-site detectors, such as
\gls{ligo}, are essential for false-positive rejection.  Coordinated telescope
observations is logistically difficult but would prove invaluable in reporting
detections.  Telescopes do not need to be observing at the same frequency, only
observing the same approximate field of view. Detection of an event at multiple
bands is strong evidence for an astrophysical source. Though, multi-frequency
observation campaigns of FRB\,121101 have shown that detections can be
band-limited \citep{2017ApJ...850...76L}.

Localization with an interferometric array produces similar results to a
multi-site observation. \gls{rfi} local to the site will still appear but it can
be better localized, and possibly determined to be in the near field. \gls{rfi}
internal to a single element would not correlate with other elements and would
not appear as a false-positive detection.  \gls{vlbi} detections provide the
highest confidence as the source is localized with multiple geographically
isolated telescopes.

Detection of \glspl{frb} at multiple frequencies not only adds to the
scientific understanding of the sources, but also helps to verify that they are
astrophysical.  Due to historical development of receivers for pulsar searches,
most \glspl{frb} have been detected at L-band frequencies. FRB\,110523 detected
with the GBT and the multiple \glspl{frb} detected with UTMOST occurred at UHF
frequencies.  Only FRB\,121102 has been detected above L-band
\citep{2017ApJ...850...76L,atel10675,2018Natur.553..182M,0004-637X-863-2-150}.
Such wide bands show the pulse structure goes beyond the bandwidth of known
sources of \gls{rfi} (e.g. modulated radar). 

Multiple \glspl{frb} (e.g. FRB\,121102, FRB\,140514, FRB\,180301) have now been
detected which are not broad-band, possibly due to the intrinsic emission of
the source, lensing, or scintillation. It is likely that additional  narrow-band
\glspl{frb} exist in past survey data but do not pass the detection threshold. A
pulse search over sub-bands could reveal further detections as the \gls{snr}
would increase.

An ideal \gls{frb} search experiment would consist of at least three stations
geographically separated to reduce common \gls{rfi} and allow for localization.
Each station would consist of an aperture array or a compact array of
small-diameter/wide field of view elements. The effective bandwidth would be
sufficient to accurately determine a dispersion relation.  Coherent beams would
be formed across the primary beam. Each station would observe the same field.
Detections at multiple sites would then trigger a capture of the complex
voltages in a transient buffer which could then be correlated after the fact. 

Though this idealized experiment does not exist, it is currently possible to
make experiments that are close to it. LOFAR stations, though very low-frequency
compared to previous detections, can be configured to operate in a similar mode.
Dense arrays such as UTMOST \citep{2017MNRAS.468.3746C}, which has detected
numerous \glspl{frb}, and CHIME \citep{Chime2018} provide significant
instantaneous sky coverage but limited spatial resolution. The planned build-out
of the Molonglo array to UTMOST-2D will provide further localization.
Additional outrigger elements such as those planned with HERA
\citep{2017PASP..129d5001D} would provide an improvement in spatial resolution.
TRAPUM\footnote{http://www.trapum.org/}, one of the MeerKAT legacy science
projects,  and MeerTRAP, a commensal transient search program on MeerKAT, will
perform single pulse surveys in beamformed data and capture complex voltage data
for post-detection localization.  Telescopes that are part of \gls{vlbi}
networks use the same back-ends and contain complex voltage recorders, can be
used \citep{2018arXiv180401904W}. Only a few telescopes in the network that can
have the same pointing are needed at a time for localization.

\section{Conclusion}

From an observational point of view, \glspl{frb} are unique astronomical sources
as, so far, no follow-up observations have been able to verify a source using a
different telescope or observing frequency, except in the case of FRB\,121102
which is known to repeat. Thus, it is necessary to provide evidence for an
astrophysical origin when reporting a detection.

In Section \ref{sec:verify_crit} we have presented a set of criteria which can
be used to verify an \gls{frb} detection relative to a prototypical \gls{frb}
model and observational data. We have shown which of the criteria are more
essential to verification than others.  For example, the strongest evidence for
an event being of astrophysical origin is a multi-site detection.  In
combination these criteria become stronger evidence for the origin of a detected
\gls{frb}. Effort should be made to test these criteria using the available data
when reporting a detection. The set of criteria and methods for applying them
laid out here should ideally be applied to all new FRB detections. Publishing a
table similar to Figure \ref{fig:heat_map} provides comprehensive information
for assessing new detections.

As \gls{frb} surveys continue to increase in sensitivity, sampled
parameter space, and time, there will be an increase in the number of
false-positive detections, even as rejection models are improved.
Differentiating between true, astrophysical \glspl{frb} and false-positive
events will become more difficult.  Further, it could be that most astrophysical
\glspl{frb} are not prototypical as it is defined now, such as in the case of
FRB\,121102. Which indicate that some of the criteria must be relaxed,
making the differentiation more difficult still.  This should warrant the drive
towards interferometric and multi-site observations, complex-voltage data
recorders, sub-band pulse searches, and standard reporting of the observing
system.

Automation of some of the verification tests can help to formalize criteria into
metrics.  Machine learning-based models are routinely employed to detect and
classify candidate detections. \cite{Zhang2018} used simulated pulses to train a
deep neural network detection model.  \cite{Connor2018} built a classifier model
based on the dynamic spectrum, pulse profile, and DM-trial space of a detection.
This model was similarly trained with simulated pulses.  UTMOST \gls{frb}
detections \citep{2018MNRAS.478.1209F} are automated through the use of a
classifier model which takes into account multi-beam detections to reduce the
number of false-positive detections due to local \gls{rfi}. The ALFABURST system
\citep{2018MNRAS.474.3847F} which detected the event in Section
\ref{sec:D20161204} uses a similar classifier model.  Inherent to these models
is a definition of a prototype pulse. They provide the initial assessment of the
detection, which can then be followed by testing the more complex criteria which
still require expert analysis.

Detection reporting incurs an economic cost in observing time, resources, and
human effort. Though, there is a scientific cost to delayed reporting as
follow-up observations could detect further pulses or multi-wavelength
observations could reveal an unknown counterpart to the radio pulse. An initial
detection will not be able to be verified against many of the criteria presented
here. But basic tests such as \gls{snr}, dispersion relation, and telescope
state can be automated to determine an initial detection `importance' in a
VO-Event trigger \citep{2017arXiv171008155P}. This importance factor can then be
adjusted accordingly as more criteria are tested.

Reporting of false-positive events, even if the source is not explained, helps
to improve the robustness of search pipelines against systematics and \gls{rfi}.
Reporting these events also helps to improve the case for \glspl{frb} being of
astrophysical origin, just as the explanation for Perytons
\citep{2015MNRAS.451.3933P} removed doubt about detections using Parkes. An
attempt should be made to make the raw data, either unprocessed filterbank or
complex-voltage data, available to be used in independent verification and
testing on search pipelines.

Jupyter notebooks of the verification tests and terrestrial \glspl{frb} are
hosted on our public git
repository\footnote{https://github.com/griffinfoster/terrestrial-frb-letter}.

\section*{Acknowledgements}

We thank Simon Johnston for his valuable comments.  A.K. and G.F. would like to
thank the Leverhulme Trust for supporting this work.  G.F. and D.C.P.
acknowledge support from the Breakthrough Listen Initiative.  Breakthrough
Listen is managed by the Breakthrough Initiatives, sponsored by the Breakthrough
Prize Foundation.  ALFABURST activities are supported by a National Science
Foundation (NSF) award AST-1616042. M.P.S. and D.R.L. acknowledge support from
NSF RII Track I award number OIA--1458952. K.R. acknowledges funding from the
European Research Council grant under the European Union's Horizon 2020 research
and innovation programme (grant agreement No. 694745).

\bibliographystyle{mnras}
\bibliography{frb-detections} 

\bsp	
\label{lastpage}
\end{document}